\documentclass[acmlarge]{acmart}

\usepackage{pifont}
\usepackage{makecell}
\usepackage{booktabs}
\usepackage{multirow}
\usepackage{graphicx}
\usepackage{amsmath}
\usepackage{enumitem}
\usepackage{subcaption}
\usepackage{ragged2e}
\usepackage{adjustbox}
\usepackage{threeparttable}
\usepackage{wrapfig}
\usepackage{algorithm}
\usepackage{algpseudocode}
\usepackage[most,skins,theorems]{tcolorbox} 
\tcbset{
  aibox/.style={
    width=\linewidth,
    top=8pt,
    bottom=4pt,
    colback=blue!6!white,
    colbacktitle=black,
    colframe=black,
    enhanced,
    center,
    attach boxed title to top left={yshift=-0.1in,xshift=0.15in},
    boxed title style={boxrule=0pt,colframe=white,},
    fonttitle=\bfseries,
  }
}
\newtcolorbox{AIbox}[2][]{aibox,title=#2,#1}

\makeatletter
\renewcommand{\@mkauthorsaddresses}{%
  \ifnum\num@authors>1\relax
    Authors' %
  \else
    Author's %
  \fi
  Contact Information:
  \bgroup
  \def\streetaddress##1{\ClassWarning{\@classname}{ACM no longer collects
    authors' postal addresses. I am ignoring your street address}%
    \unskip\ignorespaces}%
  \def\postcode##1{\ClassWarning{\@classname}{ACM no longer collects
    authors' postal addresses. I am ignoring your postal code}%
    \unskip\ignorespaces}%
  \def\position##1{\unskip\ignorespaces}%
  \gdef\@ACM@institution@separator{, }%
  \def\institution##1{\unskip\@ACM@institution@separator ##1%
    \gdef\@ACM@institution@separator{ and }}%
  \def\city##1{\unskip, ##1}%
  \def\state##1{\unskip, ##1}%
  \renewcommand\department[2][0]{\unskip\@addpunct, ##2}%
  \def\country##1{\unskip, ##1}%
  \def\and{\unskip; \gdef\@ACM@institution@separator{, }}%
  \def\@author##1{\mbox{##1}}%
  \def\@correspondingauthormark{%
      \unskip\space\mbox{(corresponding author)}}%
  \def\email##1##2{%
    \unskip, \href{mailto:##2}{\textcolor{black}{##2}}}%
  \addresses
  \egroup}
\makeatother

\AtBeginDocument{%
  }

\setcopyright{cc}
\copyrightyear{2026}
\setcctype{by}
\acmJournal{IMWUT}
\acmYear{2026} \acmVolume{10} \acmNumber{3} \acmArticle{182}
\acmMonth{9} \acmDOI{10.1145/3831658}

\begin{document}

\title{Act2Intention: A Benchmark For Developing Active Mobile Agents Through Inferring User Intention from GUI Actions}


\author{Xiaokai Yan}
\orcid{0000-0002-8139-2231}
\affiliation{%
  \institution{Northwestern Polytechnical University}
  \country{China}}
\email{kely@mail.nwpu.edu.cn}

\author{Jingtao Ding}
\correspondingauthor
\orcid{0000-0001-7985-6263}
\authornote{Corresponding author.}
\affiliation{%
  \institution{Tsinghua University}
  \country{China}}
\email{dingjt15@tsinghua.org.cn}

\author{Yong Li}
\orcid{0000-0001-5617-1659}
\affiliation{%
  \institution{Tsinghua University}
  \country{China}}

\author{Zhiwen Yu}
\orcid{0000-0002-9905-3238}
\affiliation{%
  \institution{Northwestern Polytechnical University}
  \country{China}}
\email{zhiwenyu@nwpu.edu.cn}

\renewcommand{\shortauthors}{Yan et al.}

\begin{abstract}
Mobile GUI Agents powered by multimodal large language models (MLLMs) show promise in human-computer intelligence. However, current research primarily focuses on reactive task execution while lacking a comprehensive ``understanding-prediction-execution'' process for user intentions, which are the core requirements of active agents. In this paper, we propose the Act2Intention framework that builds an active mobile agent by integrating understanding, predicting user intentions, and executing decisions. Firstly, we constructed the Act2Intention Bench through collection and validated generation, comprising 72,511 intentions and over 700,000 actions across 52 apps, thereby establishing the first benchmark for evaluating proactive agents via continuous ``intention-actions'' trajectories. We further develop the Act2Intention Agent, achieving proactive services through Proactive-oriented Intention Understanding, Personalized Proactive Intention Prediction, and Experience-guided Intention Execution. 
Experimental results show that \textcolor{black}{supervised fine-tuning on Act2Intention Bench yields absolute improvements of +32.0 Acc-S, +10.25 Acc-S, and +6.9 SSR points over non-fine-tuned counterparts under the same agent framework} for intention understanding, prediction, and execution, respectively.
This success underscores the necessity and value of the Act2Intention Bench, which establishes a standardized platform for developing and evaluating proactive agents and consequently paves the way for research on intention-driven human–computer interaction. To facilitate further research, we open-source our code and Act2Intention Bench at: \href{https://github.com/npuNancy/Act2Intention}{npuNancy/Act2Intention}.

\end{abstract}



\received{1 November 2025}
\received[revised]{1 May 2026}
\received[accepted]{1 July 2026}

\maketitle

\begin{figure}[htb]
  \centering
  \includegraphics[width=1\linewidth]{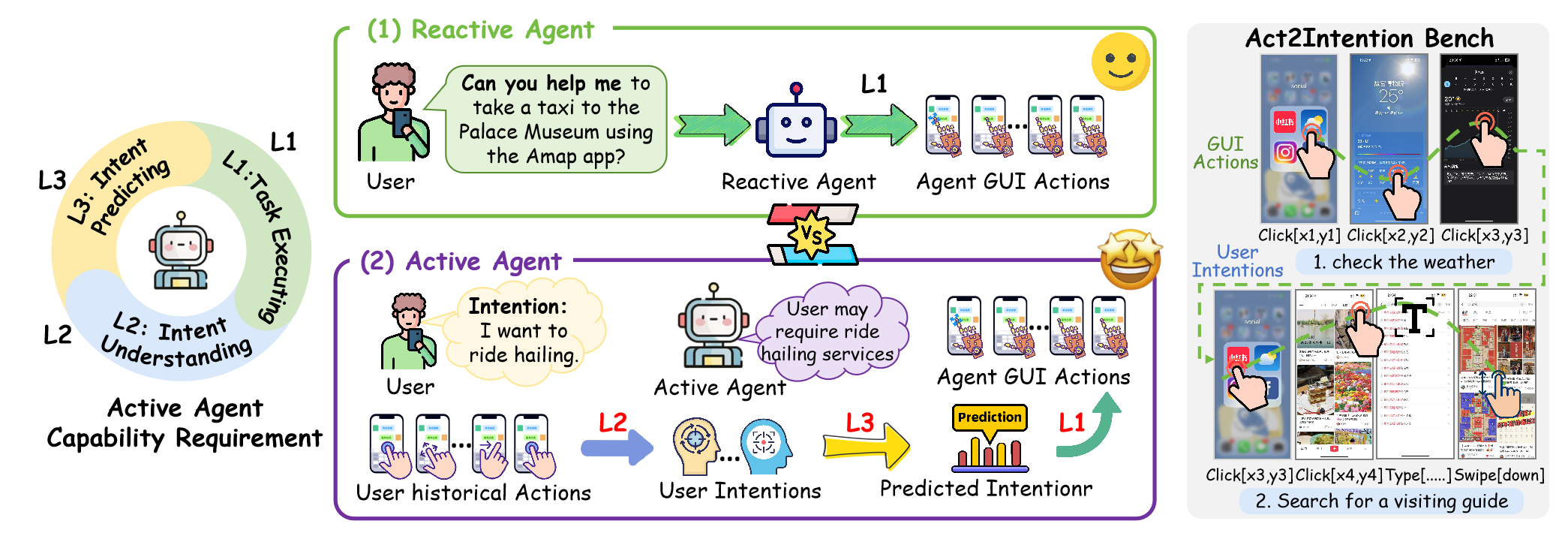}
  \caption{\textcolor{black}{\textbf{The Act2Intention Framework illustrates a shift from reactive to active agents}.} \textbf{(1) \textit{Reactive Agent}} can only passively assist until users send instructions. \textbf{(2) \textit{Active agent}} understands, predicts the user’s intention through GUI actions, and thereby provides active services (Understanding $\rightarrow$ Prediction  $\rightarrow$ Execution). A toy example of Act2Intention Bench is shown on the right.}
  \label{fig:intro}
\vspace{-4mm}
\end{figure}

\section{INTRODUCTION}

The remarkable advancement of Large Language Models (LLMs) and Visual Language Models (VLMs) is giving rise to the proliferation of autonomous agents \cite{shang2024agentsquare}. Among these, mobile Graphical User Interface (GUI) agents have evolved from rule-based systems \cite{mazumder2020flin,qian2020roscript} to LLM-powered architectures \cite{hong2024CogAgent,zhang2025TongUI}. These agents perceive environmental information by analyzing multimodal observations such as screenshots and accessibility trees, autonomously reason and execute actions via GUI interactions, to complete user instructions \cite{wang2024surveygui,shi2025surveytowards}. For instance, AutoGLM, a cross-platform GUI-Agent that spans mobile, web, and PC platforms, can emulate human cognitive processes and interaction patterns to process tasks ranging from data mining and analysis to report generation \cite{liu2024autoglm}. \textcolor{black}{Such advances may change how users interact with mobile and ubiquitous systems} \cite{qin2025UITARS}.

However, as shown in Figure~\ref{fig:intro}, current research on GUI Agents \cite{lin2024showui,lu2025UIR1,xia2025guiGUIR1}, serving as reactive actors, focuses primarily on task-executing (L1), which relies on explicit user instructions. Instead, real-world applications reveal \textcolor{black}{a practical gap}: users are often not inclined to articulate their intentions explicitly via text or voice inputs in many scenarios that require foresight and autonomous decision-making \cite{roche2016superforecasting,hohwy2013predictivemind}. 
\textcolor{black}{
Here, the \textit{user intention} denotes the underlying purpose behind a sequence of GUI actions (e.g., ``Order a no-ice Starbucks Latte via Meituan''), and a formal definition is given in Section~\ref{sec:task_define}. The specific gap addressed in this paper is continuous mobile GUI intention modeling: inferring a sequence of intention segments from raw GUI action streams, using historical intention trajectories and user personas for anticipation, and then grounding user-confirmed intentions into executable GUI actions.
Current research, such as ``Proactive Agent'' \cite{lu2024proactive}, has begun to shift the focus from reactive actor to proactive agent. However, it still lacks a cognitive architecture for understanding user intentions from raw GUI action streams. }
In summary, existing mobile agents \textcolor{black}{remain limited in their proactive service capabilities}. \textcolor{black}{To address this gap, we propose a systematic capability framework for proactive agents that includes three components: understanding $\rightarrow$ prediction $\rightarrow$ execution.}

\textcolor{black}{In this work, we present \textbf{Act2Intention}, a data-driven framework that understands and predicts user intentions by analyzing historical GUI actions, and supports user-confirmed execution to assist users in accomplishing their goals.} However, existing GUI datasets focus only on discrete task execution, lacking continuous ``intention-actions'' trajectories required for studying proactive agents. \textcolor{black}{To fill this gap, we construct Act2Intention Bench, a benchmark that captures continuous human--computer interaction flows for proactive intention modeling} (as shown in Figure~\ref{fig:intro}). Specifically, we collect real-world user ``intention-actions'' trajectories. To enhance scalability and diversity, we develop an LLM-based simulator to generate additional trajectories conditioned on diverse user personas. \textcolor{black}{Then, a two-step verification process helps improve the fidelity and diversity of the data.} The final benchmark contains 360 personas, 72,511 intentions, and over 700,000 actions across 52 mobile applications.

Based on Act2Intention Bench, we develop the Act2Intention Agent, a multi-agent framework. It consists of three specialized modules: 1) Intent Understanding, designed to understand user intention underlying GUI actions; 2) Intent Prediction, which infers potential user intentions by integrating historical intentions and their characteristics; 3) Intention Execution, leveraging historical experience to guide task execution. \textcolor{black}{To evaluate the benchmark, we fine-tune and test several open-source LLMs, including Qwen2.5-7B \cite{yang2024qwen25}, Llama3.1-8B \cite{touvron2023llama}, Deepseek-7B \cite{guo2025deepseek} and Mistral-7B \cite{jiang2023mistral7b}, on Act2Intention Bench across the three tasks of understanding, prediction, and execution.}

In summary, our contributions are as follows:
\begin{itemize}
    \item \textcolor{black}{We propose an anticipatory mobile-agent paradigm: Understanding → Predicting → Executing. This paradigm defines a framework that (1) understands user intentions from atomic action sequences, (2) predicts future intentions from historical trajectories and personas, and (3) executes user-confirmed intentions through GUI actions.}
    \item \textcolor{black}{We construct Act2Intention Bench, a mobile benchmark with continuous multi-intention and action trajectories. It contains 360 personas, 72,511 intentions, and over 700,000 actions across 52 apps, supporting the evaluation of mobile GUI Agents under continuous and personalized intention modeling.}
    \item \textcolor{black}{To evaluate the utility of Act2Intention Bench, we develop the Act2Intention Agent, which examines the feasibility of inferring and executing intentions from raw GUI actions.}
    \item \textcolor{black}{Experimental results show that the Agent, trained on Act2Intention Bench, achieves improved performance in understanding, predicting, and executing intentions, supporting the value of Act2Intention Bench.}
\end{itemize}

\section{RELATED WORKS}

\subsection{GUI Agents}  
The powerful visual perception and reasoning ability of MLLMs have enhanced the evolution of GUI Agents from rule-based automation to highly automated and generalized systems \cite{wang2024surveygui,zhang2024GUIsurveyLarge}. LLM-based GUI Agents typically require five capabilities: perception, planning, execution, reflection, and memory \cite{nguyen2024surveygui,shi2025surveytowards}. Some research, such as MobileAgent \cite{wang2024MobileAgentV2,wang2025MobileAgentE,ye2025MobileAgentV3}, employs multi-agent frameworks with precisely designed workflows independent of Supervised Fine-Tuning (SFT) \cite{yang2024aria,wen2024autodroidV1,wen2024autodroidV2,zhang2023AppAgentV1,li2024AppAgentV2}. 
Another branch of research, such as CogAgent \cite{hong2024CogAgent} and AutoGlm \cite{liu2024autoglm}, focuses on \textcolor{black}{improving} a single agent's performance through SFT \cite{xu2024androidlab,lin2024showui,zhang2025TongUI,sun2024OsGenesis,huang2025spiritsight,xu2024Aguvis,chen2025atlas}. Furthermore, UI-TARS \cite{qin2025UITARS} and Wepo \cite{liu2025wepo} involve Direct Preference Optimization (DPO) behind SFT to maximize data utility. Recent studies focus on reducing dependence on large-scale datasets \cite{liu2025infiguiR1,lu2025UIR1}. Agents like GUI-R1 \cite{xia2025guiGUIR1} replace supervision with reinforcement learning reward models, achieving better performance with minimal expert data.

\begin{table}[htbp]
\caption{Dataset comparison.}
\label{tab:Dataset_comparison}
\begin{tabular}{lllll}
\toprule
Dataset        & Episodes/Events & Platform        & Avg.Steps & Continuity? \\ \midrule
AndroidControl & 15283    & Mobile         & 5.5   & no         \\
AITW           & 715,142  & Mobile         & 6.5   & no         \\
AITZ           & 2,504    & Mobile         & 7.5   & no         \\
AMEX           & 8,000    & Mobile         & 12.8  & no         \\
GUI-Odyssey    & 7,735    & Mobile         & 15.4  & no         \\    \midrule
ProactiveBench & 6,790    & Windows        & -     & yes         \\  
\textcolor{black}{ProAgentBench}  & \textcolor{black}{28,528}   & \textcolor{black}{Windows}        & \textcolor{black}{-}     & \textcolor{black}{yes}        \\  \midrule
Act2Intention  & 72,511   & Mobile         & 13.2  & yes       \\ \bottomrule
\end{tabular}
\end{table}

\subsection{GUI Datasets} 
\textcolor{black}{GUI datasets are an important resource for developing GUI Agents.} Currently, widely used mobile GUI datasets include Android in the Wild (AiTW) \cite{rawles2023AiTW}, Android Control \cite{li2024AndroidControl}, AMEX \cite{chai2024Amex}, GUI Odyssey \cite{lu2024guiodyssey}, AiTZ \cite{zhang2024AITZ}, and others. These datasets typically contain three core components: 1) user instructions that provide overall objectives for agents, 2) environmental information comprising screenshots or UI element trees, and 3) task trajectories detailing action sequences for task completion \cite{nguyen2024surveygui}. \textcolor{black}{However, these datasets are less aligned with our target setting.} First, their task instructions are discrete rather than temporally continuous, \textcolor{black}{making them less suitable} for modeling human continuous intentions. Second, these datasets do not contain contextual information related to user intent, such as time, scene, etc. Consequently, a research gap remains in developing a benchmark for evaluating proactive agents capable of intent understanding and intent prediction. The detailed comparison between our Act2Intention Bench and other datasets is presented in Table~\ref{tab:Dataset_comparison}.

\subsection{Active Agents} 
Recent research has introduced active agents into dialogue systems, so that the system can be aware of long-term conversational goals and proactively guide dialogues toward the goals \cite{qian2024tell,Deng2024TowardsHumanCentered,zhang2024ask}. 
Beyond dialogue systems, active agents have also been applied to embodied tasks \cite{Zhang2024ProAgent,du2024CHAIC,suninteractgen,cao2024smarthelp,ding2024AtomBot}. These active bots identify potential intentions and assist humans in completing various operations based on observed actions and emotions without explicit human instructions. 

In the context of mobile and ubiquitous systems, several studies focus on predicting opportune moments for proactive interactions \cite{TowardProactive,ExploringUser,WatchingScreen}. For example, \cite{Beyond2017Pielot} developed a model that predicts when users are open to engaging with notifications, achieving significantly higher success rates by incorporating phone-use behavior. Similarly, \cite{HelloThere} and \cite{UnderstandingUser} explored interruptibility in smart speaker interactions, identifying key contextual factors such as user mood, activity, and social presence that influence the appropriateness of proactive engagements. \textcolor{black}{Recent TPCI studies have also explored related directions in context-aware agents, multimodal intent understanding, and mobile notification management \cite{JadoonYS24,WangFW24,KamalRNCM24,WeiLRHRFP26}. These works further motivate our focus on intention inference in mobile and pervasive computing.} Further extending proactive support, Yang et al. \cite{SocialMind} introduced an LLM-based AR system that provides in-situ social assistance by perceiving multi-modal cues and proactively generating suggestions during live conversations.

In GUI Agents, Zhao et al. \cite{zhao2025appagentPro} proposed AppAgent-Pro, which infers potential user needs after receiving a user instruction. Lu et al. \cite{lu2024proactive} further introduced Proactive Agent, which predicts and initiates tasks without explicit human instructions. \textcolor{black}{They also constructed ProactiveBench with about 6.8k events and improved agent proactivity through fine-tuning.} 
\textcolor{black}{More recently, Tang et al. \cite{tang2026ProAgentBench} proposed ProAgentBench, which studies proactive assistance in computer-use scenarios and evaluates agents' ability to infer and provide helpful assistance from user activity context. These works are closely related to ours, but Act2Intention focuses on a different setting: continuous mobile GUI intention modeling. Instead of an event-level task proposal, Act2Intention studies how agents infer multiple intention segments from low-level GUI actions, predict the next intention with behavior-derived personas, and execute the predicted intention on mobile devices.} \textcolor{black}{Thus, our main focus is mobile GUI-based continuous intention--action trajectories, persona-conditioned prediction, and an integrated understanding--prediction--execution benchmark.}

\section{TASK DEFINITION}\label{sec:task_define}

\textcolor{black}{
We organize the benchmark into three agent-level tasks: Intention Understanding, Intention Prediction, and Intention Execution.}

\textcolor{black}{
In the context of mobile GUI interaction, a \textit{user intention} refers to a coherent, self-contained objective that motivates a contiguous segment of GUI actions within a single mobile usage session. Each intention is expressed as a natural-language description specifying the application involved, the operation performed, and the purpose of that operation, such as ``Order a no-ice Starbucks Latte via Meituan''. Each intention can be directly mapped to a short, executable action trajectory on the device.
Here, \textit{action} (e.g., ``click'' or ``type'') is an atomic, device-level GUI operation, and an intention aggregates multiple such actions into a semantically coherent unit.
A single usage session may contain multiple sequential intentions (e.g., searching for a restaurant, then booking a ride), which together reflect the user's broader activity context. In addition, the user's intention is continuous. The boundaries between consecutive intentions are typically marked by application switches, significant shifts in operational purpose, or natural breakpoints in the interaction flow. Figure~\ref{fig:intro} illustrates this distinction with a toy example showing how a raw action stream is segmented and labeled into discrete intentions.
}


\textcolor{black}{To avoid a mismatch between theoretical formalism and the implemented benchmark, we define Act2Intention using direct sequence notation. Let $a_t$ denote the atomic GUI action at step $t$, and let $o_t$ denote the observable GUI context at that step, such as the screenshot. For the $i$-th user intention $I_i$, its action trajectory is defined as:
\begin{equation}
\label{eq:Intention_i}
\tau_i = \{(o_{p_i},a_{p_i}),(o_{p_i+1},a_{p_i+1}),\cdots,(o_{q_i},a_{q_i})\}, 
\end{equation}
where $p_i$ and $q_i$ \textcolor{black}{represent} the start and end indices of \textcolor{black}{the $i$-th} intention segment. The action description is $d_t$, and the corresponding description trajectory is $\tau_i^{des}=\{d_{p_i},d_{p_i+1},\cdots,d_{q_i}\}$. The segmentation of a continuous action stream into $m$ intention groups is represented as $\mathcal{G}=\{(p_i,q_i)\}_{i=1}^{m}$, and each group is associated with a natural-language intention label $I_i$.}


\begin{table}[htbp]
\centering
\caption{\textcolor{black}{Summary of key notations used in this paper.}}
\label{tab:notation_summary}
\begin{tabular}{ll|ll}

    \toprule
    \textbf{Notation} & \textbf{Meaning}                        & \textbf{Notation}       & \textbf{Meaning}                                   \\ \midrule
    \textcolor{black}{$a_t$ }            & \textcolor{black}{Atomic GUI action at step $t$}           & \textcolor{black}{$o_t$}                   & \textcolor{black}{GUI observation (screenshot + context) at step $t$} \\
    \textcolor{black}{$d_t$ }            & \textcolor{black}{Natural-language description of $a_t$}   & \textcolor{black}{$I_i$}                   & \textcolor{black}{The $i$-th user intention}                          \\
    \textcolor{black}{$\tau_i$ }         & \textcolor{black}{Action trajectory for intention $I_i$}   & \textcolor{black}{$\hat{I}_{t+1}$}         & \textcolor{black}{Predicted next intention }                          \\
    \textcolor{black}{$\tau_i^{des}$ }   & \textcolor{black}{Action description trajectory for $I_i$} & \textcolor{black}{$p_i,\, q_i$}            & \textcolor{black}{Start/end indices of the $i$-th intention}          \\
    \textcolor{black}{$T$ }              & \textcolor{black}{Current timestamp }                      & \textcolor{black}{$m$}                     & \textcolor{black}{Total number of intentions in a session}            \\
    \textcolor{black}{$P$ }              & \textcolor{black}{User persona}                            & \textcolor{black}{$f_\phi,f_\theta,f_\pi$} & \textcolor{black}{Models for understanding, prediction, execution }  \\ \bottomrule

\end{tabular}
\end{table}

\subsection{Intention Understanding}
In the proposed Act2Intention framework, the first task involves interpreting the user's operational trajectory—specifically, parsing the underlying user intentions $I_{1:m}$ from the observed action sequence $\tau_{1:m}$.
\textcolor{black}{This task jointly includes session segmentation and semantic intention description,} defined as:

\begin{equation}
\label{eq:IntentParse}
I_{1:m} = f_{\phi}(\tau_{1:m}^{des}) = f_{\phi}(\{ d_{p_i},d_{p_i+1},\cdot\cdot,d_{q_i}\}_{i=1}^{m}).
\end{equation}

\subsection{Intention Prediction}
Then the Agent proactively infers the user's intention to propose a task suggestion that the user may perform. Specifically, the agent predicts the potential intention $\hat{I}_{t+1}$ based on historical intentions $I_{1:t}$, current time $T$, and user persona $P$, as in the equation:

\begin{equation}
\label{eq:IntentPredict}
\hat{I}_{t+1} = f_{\theta}(I_{1:t}, T, P).
\end{equation}

\subsection{Intention Execution}
Finally, the Agent executes the user-confirmed intention on the mobile device. Based on the explicit task (intent) $I$, the Agent's historical action trajectory $h_t$, and the current observation $o_t$, it engages in iterative decision-making to determine the next action $a_{t+1}$. In addition, we integrate user persona $P$ and related action trajectories $\tau^*$ into the decision-making process. This can be formalized as:
\begin{equation}
\label{eq:IntentExecute}
a_{t+1} = f_{\pi}(I, \textcolor{black}{{h}_{t-1}}, o_t, \tau^*),
\end{equation}
where $f_\pi$ is a pre-trained GUI Agent, and ${h}_{t-1}=\{\{o_i,a_i\}_{i=1}^{t-1}\}$ is the observation-action history up to time $t-1$. $\tau^*$ is an action trajectory corresponding to similar tasks retrieved from the user's historical intention, which is provided as operating knowledge to the agent for decision-making.

\section{BENCH CONSTRUCTION}

In light of this identified research gap, we propose Act2Intention Bench, \textcolor{black}{a benchmark for studying intent understanding, intent prediction, and personalized intent execution.}
\textcolor{black}{
The Act2Intention Bench, which includes continuous ``intention-actions'' trajectories, was constructed through a pipeline that starts from real mobile interaction logs, derives behavior-based personas from these logs, and augments the benchmark by generating additional intention and action trajectories conditioned on either real or generated personas.
}

\textcolor{black}{
Specifically, it has three subsets: Act2Intention-RR, Act2Intention-RG and Act2Intention-GG, where \textit{\textbf{``RR''}} represents the ``Real-persona-to-Real-trajectory'', \textit{\textbf{``RG''}} represents the ``Real-persona-to-Generated-trajectory'', and \textit{\textbf{``GG''}} represents the ``Generated-persona-to-Generated-trajectory''. Specifically, the RR subset preserves real user action trajectories and pairs them with personas inferred from the same users' historical behaviors, the RG subset uses real behavior-derived personas to generate additional intention-action trajectories, and the GG subset uses generated personas to further expand behavioral diversity. 
}

\subsection{Raw Data Collection}
The real-world dataset used in this work was provided by a major smartphone manufacturer, which collected phone usage data from recruited participants. When users interact with their mobile devices, various types of operational logs were generated, desensitized, and reported with explicit user consent\footnote{The data collection process has been reviewed and approved by the company's ethics review board.}. After anonymization and security screening, the smartphone manufacturer shared the processed dataset with our research team for user behavior research. The dataset covers mobile usage logs recorded between March 1, 2024 and April 29, 2024, contributed by 90 anonymous participants. Participants were explicitly informed about the study’s purpose of intention inference, and were made aware of potential privacy risks. However, due to confidentiality agreements and privacy protection policies, we were not granted access to demographic information about participants (e.g., age, gender, or technical background), the recruitment process, withdrawal mechanism, or compensation details. Furthermore, the released dataset has been processed with safety filters to ensure that there is no harmful content or private information in our dataset.

\textcolor{black}{The raw data consists of anonymized mobile interaction logs and user-provided intention descriptions. Each log entry records the timestamp, foreground app/activity information, GUI observation, action type, action parameters (e.g., click coordinates, input text, swipe direction, or pressed key), and the short intention statement filled in by the participant when an app-switching event was detected. The detailed collection procedure is illustrated in Appendix~\ref{appendix:data_collection} and Figure~\ref{fig:data_collection}. Here, the raw action metadata follows five high-frequency operation types: CLICK, LONG\_CLICK, TYPE, SWIPE, and PRESS, as shown in Table~\ref{tab:actions_Description}. These fields provide both the observable action evidence and the user-stated intention information for constructing real intention--action trajectories.}

\begin{figure}[t]
  \centering
  \includegraphics[width=1.0\linewidth]{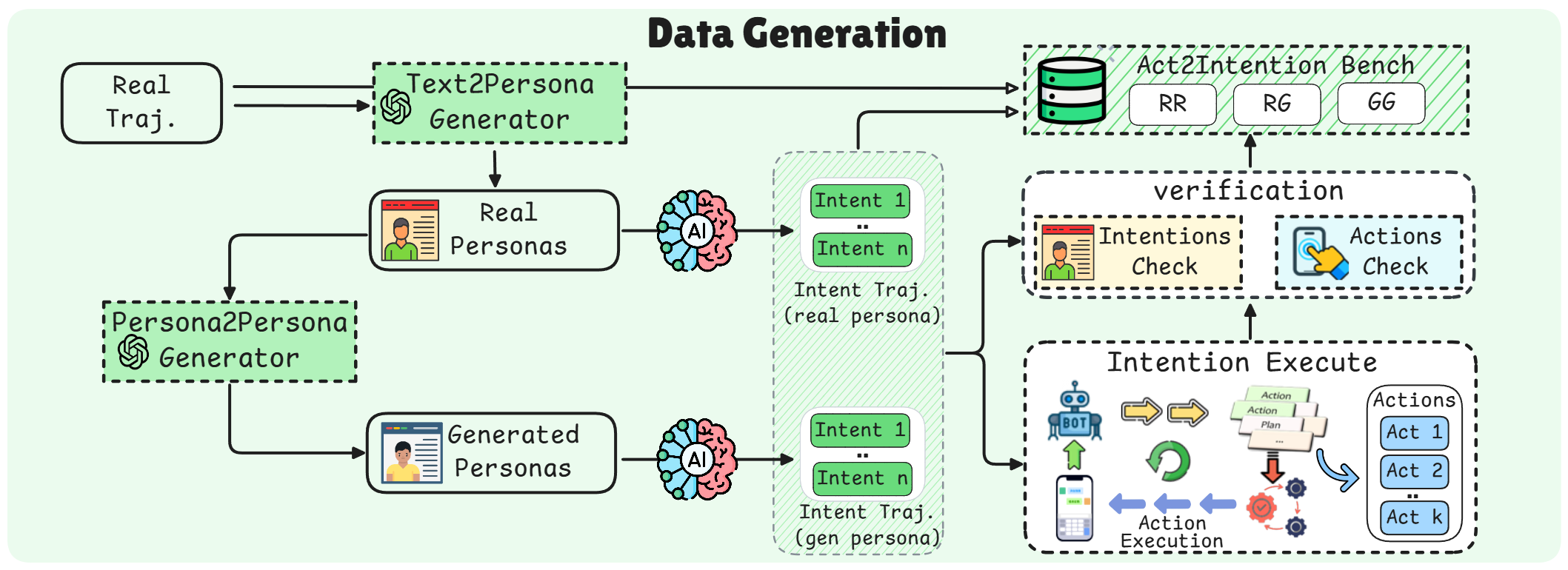}
  \caption{\textbf{Overview of bench construction}. We generate trajectory data using both real and synthesized personas with LLMs and GUI Agents, followed by dual verification.}
  \label{fig:bench_construct}
\end{figure}

\vspace{-3mm}
\subsection{Trajectory Construction and Augmentation}

\textcolor{black}{
Based on the raw logs, we construct benchmark trajectories in two complementary ways: first, we organize real user-stated ``intention-actions'' segments collected from mobile usage sessions to form the RR subset. Second, we synthesize additional ``intention-actions'' trajectories conditioned on real or generated personas to form the RG and GG subsets.
Each example in Act2Intention Bench is represented as a continuous sequence of $N$ intention segments, i.e., $E = \tau_{1:N}  = \{\tau_{1},\cdots,\tau_{N}\}$. Each intention segment consists of timestamps $T$, contextual information $C$, state observations $o$, and action sequences $a$ required to achieve that intention, i.e., $\tau_i=\{ T_i,C_i, (o_{p_i}, a_{p_i}), (o_{p_i+1}, a_{p_i+1}), \cdots, (o_{q_i}, a_{q_i}) \}$, where $p_i$ and $q_i$ represent the start and end indices of the \textit{i}-th intention segment.}

\subsubsection{Intention Generation}
We first prepared several foundational resources to assist LLMs in generating synthetic data, including user personas, scenarios, and app collections. \textcolor{black}{In Act2Intention, a persona refers to a behavior-derived profile summarized from historical intention trajectories.} Specifically, following the methodology of Tao et al.~\cite{ge2024scaling}, we generated corresponding \textcolor{black}{behavior-derived personas} based on intention trajectories collected from 90 participants by using the Deepseek R1 model with a Text-to-Persona approach. 
Based on these generated personas, we further employed a Persona-to-Persona method to create an additional 100 personas, which enhances diversity. Subsequently, we generated a set of scenarios to provide sufficient contextual information for subsequent user intention generation. These scenario descriptions encompass time, locations, and optional app collections. 

Next, we adopted an iterative approach to generate intention trajectory data. The LLMs used the user personas (including real and generated) and scenarios to synthesize initial intention trajectories. Subsequently, the LLMs were instructed to iteratively generate extended trajectory chains by incorporating previously generated intention trajectories as contextual input. 
\textcolor{black}{Each generated intention must contain a specific timestamp, a concrete app name, and a concise user intention. Consecutive intentions are required to follow a coherent temporal and behavioral flow, so that the generated trajectory resembles a continuous mobile usage session rather than a set of independent tasks. In addition, each intention is constrained to describe only one brief and specific in-app objective, avoiding compound goals that combine multiple operations in a single description. We also require the generated intentions to be objective, feasible, and grounded in realistic app operations supported by the selected app collection. }

\textcolor{black}{Table~\ref{tab:generation_example} provides a concrete example of the persona-conditioned generation process, showing how a behavior-derived persona and scenario are converted into an intention sequence and then into executable GUI action trajectories.} The full prompt template is provided in Appendix~\ref{appendix:data_generation}.

\begin{table}[t]
\centering
\small
\caption{\textcolor{black}{An example of persona-conditioned trajectory generation.}}
\label{tab:generation_example}
\begin{tabular}{p{0.17\linewidth}p{0.75\linewidth}}
\toprule
\textcolor{black}{Stage} & \textcolor{black}{Example} \\
\midrule
\textcolor{black}{Persona} 
& \textcolor{black}{A weather-conscious and socially active individual who frequently checks weather conditions through MyObservatory across multiple cities. This user maintains social connections via WhatsApp and Discord, shops online at Etsy, and navigates with Google Maps. Usage is concentrated in the afternoon and morning hours, suggesting an outdoor enthusiast who plans activities around weather conditions.} \\ \midrule

\textcolor{black}{Scenario} 
& \textcolor{black}{Weekday schedule involving weather monitoring (MyObservatory), online shopping (Etsy), social messaging (WhatsApp), and nearby restaurant exploration (Google Maps).} \\ \midrule

\multirow{3}{*}{\textcolor{black}{Generated intentions}}
& \textcolor{black}{(1) 09:07, MyObservatory: Ask the Dr.\,Tin chatbot about the expected sunset time in New York.} \\
& \textcolor{black}{(2) 17:10, Etsy: Search ``coffee mugs'' priced \$5--\$20, sort by highest price, and add the third item to cart.} \\
& \textcolor{black}{(3) 17:40, Google Maps: Explore nearby restaurants rated over 4 stars and get walking directions to the nearest one.} \\ \midrule

\multirow{9}{*}{\textcolor{black}{Action trajectories}}
& \textcolor{black}{For the Etsy intention, the agent executes the following steps:} \\
& \textcolor{black}{(1) Click the search bar (\texttt{CLICK[344,217]}).} \\
& \textcolor{black}{(2) Type ``coffee mugs'' and press enter (\texttt{TYPE[coffee mugs]}, \texttt{PRESS\_ENTER}).} \\
& \textcolor{black}{(3) Set the price filter to \$5--\$20 (\texttt{CLICK[451,383]}, \texttt{TYPE[5]}, \texttt{TYPE[20]}).} \\
& \textcolor{black}{(4) Sort results by highest price (\texttt{CLICK[968,645]}, \texttt{CLICK[1217,1126]}).} \\
& \textcolor{black}{(5) Select the third non-ad item (\texttt{CLICK[692,1355]}).} \\
& \textcolor{black}{(6) Check the seller's rating (\texttt{CLICK[282,2133]}).} \\
& \textcolor{black}{(7) Click ``Add to cart'' (\texttt{CLICK[781,2561]}).} \\
& \textcolor{black}{Each step is recorded with a screenshot.} \\ \midrule

\textcolor{black}{Verification} 
& \textcolor{black}{The intention sequence is checked for persona consistency (e.g., weather checks align with the weather-conscious profile), and the action trajectory is verified by whether the terminal screenshot supports successful completion of the intended goal.} \\
\bottomrule
\end{tabular}
\end{table}

\subsubsection{Action Trajectory Generation}
Finally, we construct the complete ``intention-actions'' trajectory by generating corresponding action trajectories for each intention. Specifically, we deploy multiple Android emulator instances in parallel on the server to enhance the efficiency of action trajectory collection. The UI-TARS-1.5-7B~\cite{qin2025UITARS} model is then employed as an actuator to execute each synthesized user intention. At each timestep $t$, the actuator receives the intention and interaction history, observes the environment, and subsequently generates an action $a_{t+1} = f_{\pi}(I, {h}_{t-1}, o_t)$. Through iterative execution, this sequence of actions $\tau=\{  (o_1,a_1), (o_2,a_2), \cdots \}$ constitutes an intention-specific execution trajectory. By concatenating the execution trajectories corresponding to the intention sequence, we obtain the final ``intention-actions'' trajectory.

\subsubsection{Data Quality Check.} 
\textcolor{black}{To improve the reliability of synthetic data, we perform the following two key verification steps.} First, we verify whether the generated intention trajectory aligns with the corresponding user persona. Using Deepseek-R1, we reconstruct the user persona from synthetic intention sequences and calculate cosine semantic similarity metrics following the setting of Tao~\cite{ge2024scaling}. Synthetic sequences exhibiting similarity lower than 0.7 are filtered out, thereby preserving alignment between synthetic and collected data. Second, we validate the completeness and accuracy of action trajectories in executing their corresponding intentions. 
\textcolor{black}{
Leveraging DeepSeek-R1 to analyze trajectories based on the intention, step-level action metadata, action descriptions, and the terminal screenshot, we evaluate whether the trajectory endpoint achieves the intended goal. The validation prompt asks the evaluator to consider three criteria: completeness, correctness, and final-state plausibility. Completeness checks whether all necessary sub-goals are covered, correctness checks whether the executed actions are aligned with the given intention, and final-state plausibility checks whether the terminal screenshot provides sufficient evidence that the intended goal has been achieved. The evaluator outputs a binary judgment, and only trajectories receiving a positive judgment are retained.}

\textcolor{black}{This validation process also helps identify common failure modes in synthetic action trajectories. At the intention level, we mainly filter out trajectories that are inconsistent with the corresponding persona, overly generic, or semantically disconnected from the given scenario. At the action level, we mainly remove trajectories with incomplete operations, infeasible app states, incorrect final screens, missing critical steps, or insufficient visual evidence that the target intention has been completed. These filtered cases are excluded from the final benchmark.} Details of the quality check are provided in the Appendix~\ref{appendix:data_check}.

\begin{table}[htbp]
\centering
\caption{\textcolor{black}{Synthetic-data filtering.}}
\label{tab:synthetic_filtering}
    \begin{tabular}{lrrrr}
    \toprule
    \textbf{Stage} & \textbf{Generated} & \textbf{Retained} & \textbf{Filtered} & \textbf{Retention} \\ \midrule
    Intention filter & \textcolor{black}{73640} & \textcolor{black}{68210} & \textcolor{black}{5430} & \textcolor{black}{92.63\%} \\ 
    Action filter & \textcolor{black}{68210} & \textcolor{black}{\textbf{59362}} & \textcolor{black}{8848} & \textcolor{black}{87.03\%} \\ 
    \bottomrule
    \end{tabular}
\end{table}

\textcolor{black}{Table~\ref{tab:synthetic_filtering} reports the retention statistics of our two-step synthetic data filtering pipeline. The \textbf{Intention filter} checks whether each generated intention trajectory is consistent with its corresponding user persona. We use DeepSeek-R1 to reconstruct a persona from the synthesized intention sequence and compute the cosine semantic similarity against the original persona; sequences with similarity below 0.7 are discarded. The \textbf{Action filter} then verifies whether each executed action trajectory successfully achieves its corresponding intention. Specifically, DeepSeek-R1 is prompted with the intention description and the terminal screenshot to judge whether the trajectory endpoint realizes the intended goal. Trajectories receiving a ``False'' judgment are removed. Overall, 92.63\% of intentions and 87.03\% of action trajectories pass the two-step verification, yielding the final Act2Intention Bench.}

\subsection{Dataset Statistics and Analysis}

\begin{figure}[htbp]
	\centering
	\begin{minipage}[c]{0.48\textwidth}
		\centering
		\includegraphics[width=\textwidth]{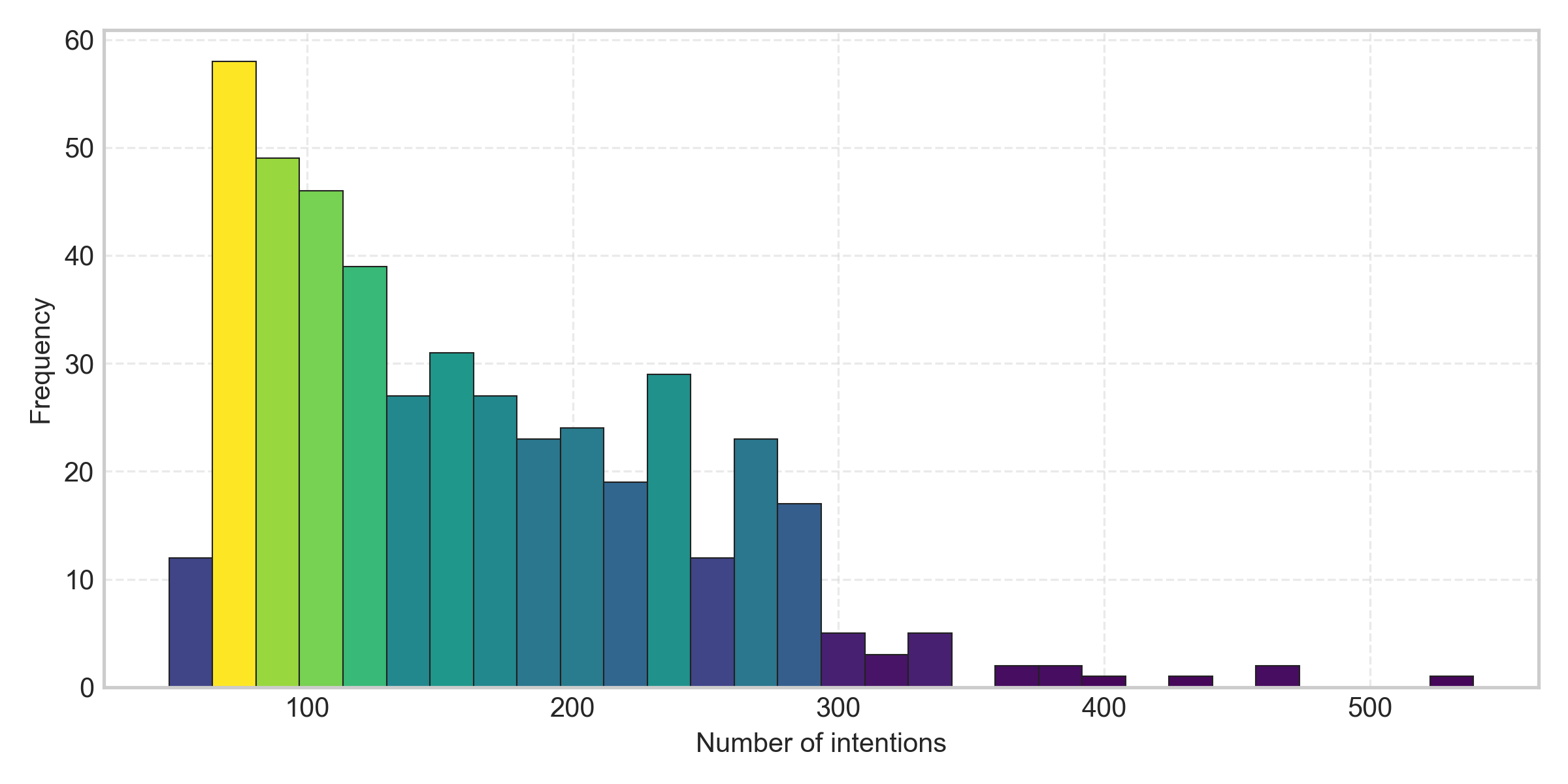}
		\subcaption{Distribution of intention length.}\label{fig:Distribution_length}
	\end{minipage} 
	\begin{minipage}[c]{0.48\textwidth}
		\centering
		\includegraphics[width=\textwidth]{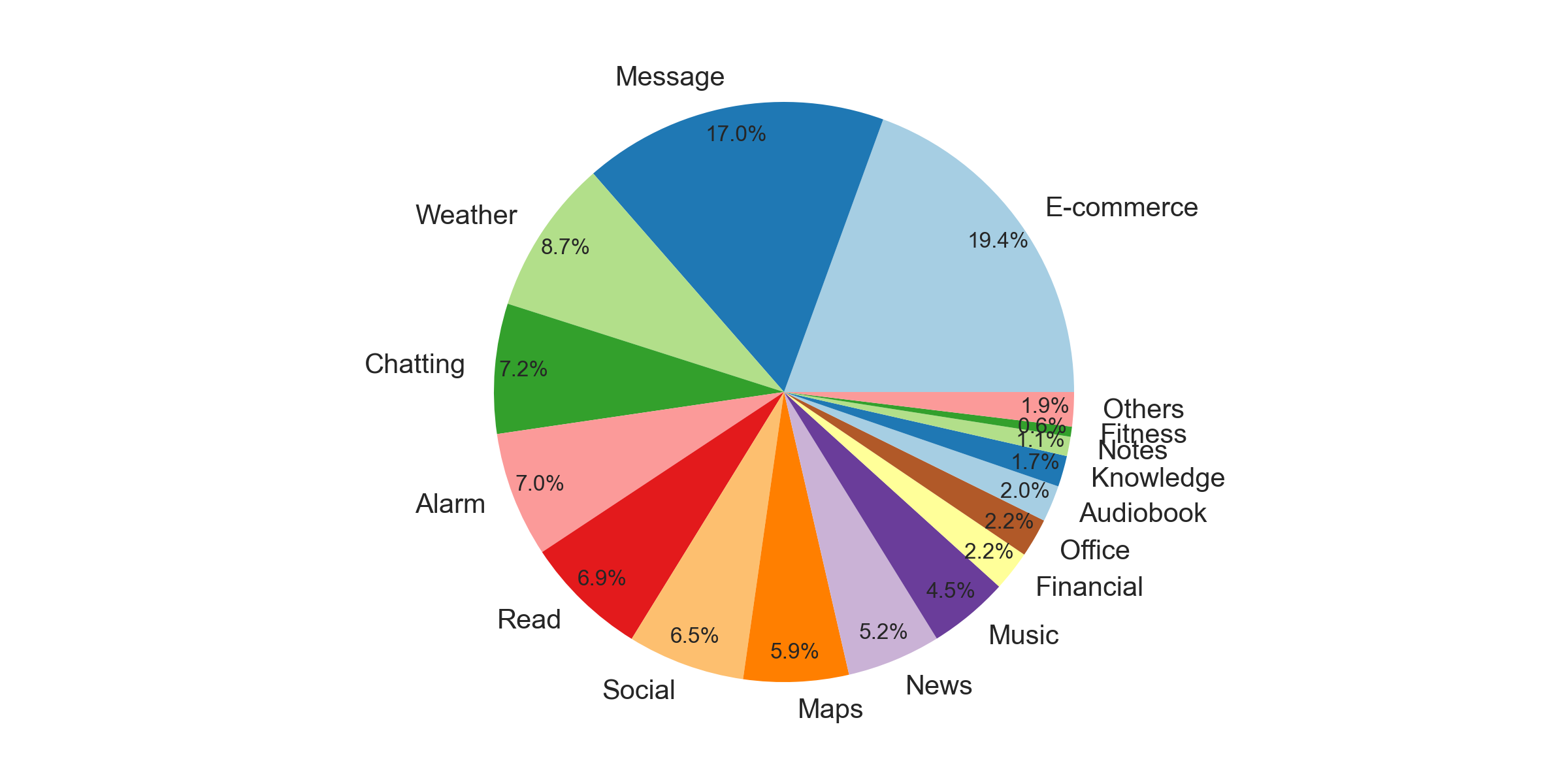}
		\subcaption{Distribution of intention categories.}\label{fig:Distribution_categories}
	\end{minipage}
    \caption{Dataset statistics and distribution.}
    \label{fig:Dataset_Statistics}
\end{figure}

Table~\ref{tab:Dataset_Statistics} shows the statistics of our dataset, including 360 personas, 72,511 intentions, and 700,000+ actions across 52 apps.
\textcolor{black}{The released trajectory schema also contains the fields ``event'' and ``domain''. For the ``event'' field, we first manually defined about 120 fine-grained intent categories based on common mobile usage scenarios, and then used gpt-4o-mini to classify each intention text into one of these categories. The ``domain'' field denotes a coarser app-level usage domain, such as communication, shopping, navigation, entertainment, or productivity, and is not used as the intent category in our evaluation. Here, Figure~\ref{fig:Dataset_Statistics} reports the distribution of the length of the intention trajectory and intention categories.}

\textcolor{black}{
\textbf{Human Realism Assessment.}
To evaluate whether synthetic intentions are perceived as realistic by human judges, we randomly sampled 30 intention segments from RR, RG, and GG, respectively. 
Each sample was evaluated by 10 annotators under a blind setting, which subset the sample came from. Details of the annotator recruitment, rating protocol, and agreement analysis are provided in Appendix~\ref{appendix:human_realism}. Annotators then rated intention realism, action achievability, and persona consistency on a 5-point Likert scale (1=clearly synthetic, 5=highly realistic). As shown in Table~\ref{tab:human_realism}, RG and GG obtain realism scores of 3.91 and 3.55, respectively, compared with 4.19 for RR. The gap indicates that synthetic trajectories are not identical to real user traces, but their high achievability and persona-consistency scores suggest that they are plausible for benchmark augmentation. \textcolor{black}{This human assessment should be interpreted as a limited sanity check, since it covers 30 intention segments per subset and 10 annotators; broader validation with larger samples and more diverse evaluators is left for future work.}
}

\begin{table}[htbp]
\centering
\caption{Statistics of Act2Intention Bench.}
\label{tab:Dataset_Statistics}
    \begin{tabular}{lllll}
    \toprule
    Subset           & Personas & Intents & P.C.Intents & Actions \\ \midrule
    Act2Intention    & 360      & 72511   & 161.14      & 705366  \\ \midrule
    \quad Act2Intention-RR & \multirow{2}{*}{90}       & 13149   & 146.10      & 134542  \\
    \quad Act2Intention-RG &          & 17678   & 196.42      & 181131  \\
    \quad Act2Intention-GG & 270      & 41684   & 154.39      & 389693  \\ \bottomrule
    \end{tabular}
\end{table}

\vspace{-3mm}

\begin{table}[htbp]
\centering
\caption{\textcolor{black}{Human realism assessment of real and synthetic trajectories.}}
\label{tab:human_realism}
    \begin{tabular}{lccc}
    \toprule
    \textbf{Subset} & \textbf{Realism} $\uparrow$ & \textbf{Achievability} $\uparrow$ & \textbf{Persona Consistency} $\uparrow$  \\ \midrule
    Act2Intention-RR & \textcolor{black}{4.19} & \textcolor{black}{4.62} & \textcolor{black}{4.30}  \\ 
    Act2Intention-RG & \textcolor{black}{3.91} & \textcolor{black}{4.00} & \textcolor{black}{4.06}  \\ 
    Act2Intention-GG & \textcolor{black}{3.55} & \textcolor{black}{3.79} & \textcolor{black}{4.01}  \\ 
    \bottomrule
    \end{tabular}
\end{table}

\subsection{Dataset Splits}

We divided the Act2Intention Bench into three subsets: a training set, an in-distribution (ID) test set, and an out-of-distribution (OOD) test set. All test data was sourced from Act2Intention-RR. The OOD test set consists of randomly selected data from 10 users in Act2Intention-RR, while the ID test set comprises the last 20\% of data from the remaining 80 users. The training set includes all remaining data (including Act2Intention-RR/RG/GG).

\section{ACT2INTENTION AGENT}

\begin{figure}[t]
  \centering
  \includegraphics[width=\linewidth]{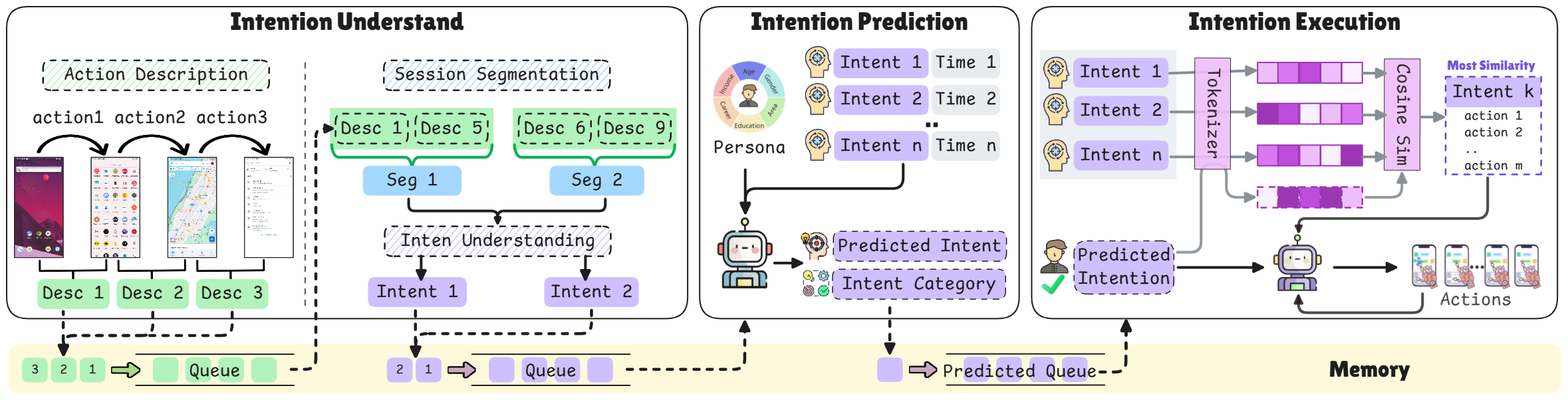}
  \caption{Overview of Act2Intention Agent: Understanding Intentions from Actions, Predicting Needs Proactively, and Executing Tasks with Experience.}
  \label{fig:Agent_Overview}
\end{figure}

Employing the data from Act2Intention Bench, we design a modular architecture, Act2Intention Agent, for proactive agent services. As defined in Section~\ref{sec:task_define}, Figure~\ref{fig:Agent_Overview} provides an overview of the Act2Intention Agent, which actively leverages raw GUI interactions to comprehend intentions, utilizes user persona context to infer latent human requirements, and ultimately provides proactive assistance when necessary.

\vspace{-5mm}

\subsection{Proactive-oriented Intention Understanding} \label{sec:Intention_Understand}
Act2Intention Agent's first step is to deduce the user's true intentions from their sequence of atomic actions with mobile devices. As shown in Figure~\ref{fig:Agent_Overview}, this process consists of three components: action description, session segmentation, and intention understanding.

\subsubsection{Action Description}
The Act2Intention Agent first employs VLMs to convert user actions into natural language descriptions, capturing the purpose and context of each action step. For each action $a_{t}$, the Act2Intention Agent integrates pre-action and post-action observations ($o_{t}$ and $o_{t+1}$) to generate action description $d_t$ in natural language, as shown in the formula in Section~\ref{sec:task_define}: $d_t = f(d_t \mid  o_t,a_t,o_{t+1})$. 

Here, the pre-action and post-action observations ($o_{t}$ and $o_{t+1}$) include both screenshots and device context information. Additionally, we highlight the action area on the pre-action screenshot. Then, the framework outputs standardized descriptions in the following format: ``\texttt{[On/In] [App/Activity], [Action], to [Purpose]}''. All generated descriptions ${d_1, d_2, \dots, d_M}$ are then stored in the Memory module. \textcolor{black}{In our framework, action descriptions serve as an intermediate semantic representation that normalizes heterogeneous GUI observations into a compact textual form for downstream intention-level reasoning.}

\subsubsection{Session Segmentation}

The raw action sequence is a continuous stream \textcolor{black}{that does not explicitly distinguish different intention segments}. The Act2Intention Agent must determine where one user intention ends and another begins. We formalize this segmentation process as:
\begin{equation}
\label{eq:segmentation}
\bigcup\limits_{i=1}^{m} \{ {p_i},\cdots,{q_i}\}=f(\{ d_1,d_2,\cdots,d_M \}), \forall i , \exists [p_i,q_i] \subseteq [1,M].
\end{equation}

Here, $M$ denotes the total number of actions in the session, $m$ is the number of distinct user intentions inferred, $[p_i, q_i]$ mark the start and end indices of the $i$-th intention segment. For instance, given an action sequence of five descriptions, the model may segment it into two intentions: 1) ``\texttt{Open Maps → Click Search Bar → Search restaurant}'', and 2) ``\texttt{Open Ride App → Book taxi}''.

\subsubsection{Intention Understanding}
After segmentation, the Agent interprets the meaning of each segment ${d_{p_i}, \dots, d_{q_i}}$ to infer the underlying intention $I_i$ according to Equation~\ref{eq:IntentParse}. Each inferred intention (e.g., “find nearby restaurants”, “book a ride”) is then stored in the Memory as structured semantic data.

\subsubsection{Training Scheme}
First, for Action Description, we utilized APIs from commercial and open-source VLMs for inference. Second, for Session Segmentation and Intention Understanding, we supervised fine-tuned LLMs to perform these tasks in a chained manner. Specifically, we constructed an SFT dataset $\mathcal{D}=\{(\mathcal{T},{Y})\}$, where $\mathcal{T}$ represents the action description sequences $\{ d_1,d_2,\cdot\cdot,d_M \}$.  The output ${Y}$ contains segmentation indices $\bigcup\limits_{i=1}^{m} \{ {p_i},\cdot\cdot,{q_i}\}$ and the corresponding intentions $I_{1:m}$. Details of the training are in the \textcolor{black}{Appendix~\ref{appendix:Act2Intention_Agent}}.

\subsection{Personalized Proactive Intention Prediction}

As shown in equation~\ref{eq:IntentPredict}, we use the historical intention sequence $I_{1:t}$, user persona $P$, and time information $T$ to predict the next potential intention $\hat{I}_{t+1}$. 
Formally, the SFT dataset is defined as $\mathcal{D}_I = {(X, Y_I)}$, where $X = (I_{1:t}, P, T)$ is the input context and $Y_I = (\hat{I}_{t+1}, c_{t+1})$ represents the predicted next intention and its category.

When the Act2Intention Agent is deployed on mobile devices, the intention prediction process runs in background mode (e.g., during system sleep). Upon device wake-up, the predicted task suggestion is displayed to the user for confirmation. \textcolor{black}{This predefined trigger is used to instantiate the prototype interface, and predicting the opportune moment for interruption is not part of the current benchmark task.}

\subsection{Experience-guided Intention Execution}

When the task suggestion predicted by the Act2Intention Agent is accepted by the user, the agent initiates experience-guided intention execution. The Act2Intention Agent \textcolor{black}{uses} similar ``intention-actions'' trajectories stored in Memory to guide the generation of more accurate operations for fulfilling the user's intent. Specifically, for each predicted intention, we first employ the all-MiniLM-L6-v2 model to convert the intention into a text embedding, thereby capturing its essential semantic information. Then, we compute the cosine similarity between this embedding and other intention embeddings within the same intention category to quantify their semantic relationships. Based on the similarity scores, the most relevant intention and its corresponding action sequence are selected as the guiding experience for execution.

Finally, following the definition in Section~\ref{sec:task_define}, we utilize GUI Agents pre-trained on GUI datasets to execute the intention by incorporating the guiding trajectory and user profile into the prompt. For each intention to be executed, the Act2Intention Agent processes it through an iterative cycle of perception, decision-making, and execution until the task is completed or the maximum step limit is reached.

\section{EXPERIMENTS} \label{Sec:Experiments}

\subsection{Experiment Setup}
\subsubsection{\textbf{Implementation Details}} 
We implemented the Act2Intention Agent using closed-source LLMs and open-source LLMs to assess their capability in intention understanding, prediction, and execution. For closed-source models, we used GPT, Claude, Gemini and Qwen-max. For open-source models, we fine-tuned Qwen-2.5-7B, Deepseek-7B, Mistral-7B, and Llama-3.1-8B on the Act2Intention Bench. Using additional LLMs would not affect the overall experimental conclusions. The details are as follows:

\textbf{Intention Understand.} As described in Section~\ref{sec:Intention_Understand}, we employed Qwen-VL-MAX for Action Description. Subsequently, we supervised fine-tuned Qwen-2.5-7B, Deepseek-7B, Mistral-7B and Llama-3.1-8B for Session Segmentation and Intention Understanding. Specifically, we constructed the training data for Intention Understanding based on the Act2Intention Bench training set, where each sample contained an intention sequence of length $n=5$. The input consisted of action description sequences corresponding to these $n$ sequences, while the output included the action sequence indices for each intention along with their natural language descriptions.

\textbf{Intention Prediction.}  Similarly, we supervised fine-tuned Qwen-2.5-7B, Deepseek-7B, Mistral-7B and Llama-3.1-8B for Intention Prediction. Using the Act2Intention Bench, we created training data where each sample's input was an intention sequence of length $m=50$, and the output was the predicted intention along with its category.

\textbf{Intention Execution.} We implemented Intention Execution by leveraging pre-trained GUI Agents: UI-TARS-2B-SFT and UI-TARS-7B-SFT~\cite{qin2025UITARS}, while incorporating additional guidance trajectories in the prompts. We evaluated the performance using the entire test set (both ID and OOD test sets). Notably, due to differences in action spaces between UI-TARS and Act2Intention Bench, we performed additional action space adaptation on the test set.

For both Intention Understanding and Intention Prediction, training used a learning rate of 1e-5 with cosine scheduling, a warmup ratio of 0.1, and a batch size of 1 over 3 training epochs. Additionally, we employed Low-Rank Adaptation (LoRA) with a rank of 64 and an alpha value of 128. All experiments were conducted on a single NVIDIA A100 GPU with 80GB VRAM.

We use the symbols $\dagger$ and $\ddagger$ to distinguish between the SFT-trained models used for Intention Understanding and Intention Prediction, respectively.

\subsubsection{\textbf{Metrics}}

\textbf{Intention Understanding.} 
We evaluated the performance of intention understanding using three complementary metrics: group accuracy (\textbf{Acc-G}), semantic accuracy (\textbf{Acc-S}), and lexical similarity (\textbf{BLEU-4}). 

\textbf{Acc-G} measures whether the action sequence is correctly segmented into corresponding intention units:
\begin{align}
\label{eq:Acc_grouping}
  \text{Acc-G} = \frac{1}{N} \sum\limits_{i=1}^{N}\mathbb{I} \left(\left( \hat{p}_i={p}_i \right) \wedge  \left( \hat{q}_i={q}_i \right) \right),
\end{align}
where $\mathbb{I}(\cdot)$ is the indicator function.

\textbf{Acc-S} assesses the semantic alignment between the predicted and ground-truth intentions by computing the cosine similarity between their text embeddings:
\begin{align}
\label{eq:Acc_semantic}
  \text{Acc-S} = \cos(\hat{I}_i,I_i) = 
  \frac{\hat{e}_i \cdot e_i}{\| \hat{e}_i \| \, \| e_i \|},
\end{align}
where $e_i$ denotes the embedding of intention $I_i$. 
This metric captures whether the model conveys a semantically equivalent intention even if the textual phrasing differs (e.g., ``book a taxi'' vs. ``order a ride''). It is widely adopted in LLM-based agent evaluation for measuring semantic fidelity beyond surface matching.

\textbf{BLEU-4}, in contrast, quantifies n-gram overlap between generated and reference intentions, emphasizing surface-level linguistic precision. It is defined as:
\begin{align}
\label{eq:BLEU}
  \text{BLEU-4} = \text{BP} \cdot \exp \left( \sum_{n=1}^{4} w_n \log p_n \right),
\end{align}
where $p_n$ denotes the modified n-gram precision, $w_n = \tfrac{1}{4}$ represents uniform weights, and $\text{BP}$ is the brevity penalty. 
Together, Acc-S and BLEU-4 offer a complementary evaluation of intention understanding.

\textbf{Intention Prediction.} 
For intention prediction, we adopt the same \textbf{Acc-S} and \textbf{BLEU-4} metrics to evaluate semantic and generative quality, and further report the category classification accuracy (\textbf{Acc-C}) to assess the correctness of intent types.

\textbf{Intention Execution.} 
We evaluate the performance of intention execution using the \textbf{Step Success Rate (SSR)}, which measures the proportion of successfully executed steps within a task. 
Formally, it is defined as:
\begin{align}
\label{eq:SSR}
\text{SSR} = \frac{1}{N_{task}} \sum_{i=1}^{N_{task}} 
\frac{1}{T_i} \sum_{t=1}^{T_i} \mathbb{I} \left( \hat{a}_{i,t} = a_{i,t}^{*} \right),
\end{align}
where $N_{task}$ denotes the number of tasks, $T_i$ represents the total number of action steps in the $i$-th task, $\hat{a}_{i,t}$ is the action executed by the agent at step $t$, and $a_{i,t}^{*}$ is the corresponding ground-truth action. 
\textcolor{black}{SSR is a conservative step-level matching metric: an executed action is counted as successful only when it matches the corresponding reference action at the same step after action-space adaptation. Therefore, functionally equivalent but different GUI paths, such as using a system back action instead of an in-app back button, may be counted as mismatches under SSR. We use SSR for fine-grained diagnosis of trajectory-level execution alignment, while the end-to-end SR metric below evaluates online task completion.}

\begin{table}[t]
\centering
\caption{Evaluation results of \textbf{Intention Understanding} and  \textbf{Intention Prediction} on the ID test set. All of the fine-tuned models achieve significant performance improvements on all metrics.}
\label{tab:Understander_Evaluation}
\vspace{-3mm}

\begin{tabular}{l|lll|lll}
    \toprule
    \multirow{2}{*}{Models} & \multicolumn{3}{|c}{Understanding} & \multicolumn{3}{|c}{Prediction} \\ \cmidrule(r){2-4} \cmidrule(r){5-7}
                        & Acc-G    & Acc-S & BLEU-4     & Acc-C  & Acc-S & BLEU-4       \\ \midrule

    \multicolumn{4}{l}{\textbf{\textit{Closed-source models}}} \\ \midrule
    GPT-3.5-turbo                & 16.31    & 0.17  & 1.91  & 42.50  & 0.29    & 17.13        \\       
    GPT-4o-mini                  & 20.66    & 0.21  & 2.31  & 38.50  & 0.31    & 18.36        \\       
    GPT-4o                       & 20.08    & 0.25  & 5.32  & 34.50  & 0.32    & 16.74        \\       
    Claude-3.5-haiku             & 24.28    & 0.24  & 4.12  & 28.50  & 0.17    & 14.27        \\       
    Claude-3.5-sonnet            & 20.00    & 0.29  & 6.58  & 15.50  & 0.24    & 16.23        \\       
    Qwen-max                     & 16.10    & 0.26  & 4.45  & 48.50  & 0.35    & 30.68        \\       
    Gemini-1.5-pro               & 68.63    & 0.32  & 8.48  & 34.50  & 0.36    & 26.86        \\ \midrule    
    \multicolumn{4}{l}{\textbf{\textit{Open-source models}}} \\ \midrule
    ProactiveAgent               & -        & -     & -     & -      & 0.22    & 17.05         \\
    Llama-3.1-8B             & 6.03                 & 0.11              & 1.41              & 37.8              & 0.33              & 24.87             \\ 
    Llama-3.1-8B-SFT & 97.31\textcolor{green}{[+91.28]}      & 0.47\textcolor{green}{[+0.36]}     & 49.83\textcolor{green}{[+48.42]}   & \textbf{54.20}\textcolor{green}{[+16.4]}  & 0.39\textcolor{green}{[+0.06]}     & \textbf{34.95}\textcolor{green}{[+10.08]} \\  \midrule
    Deepseek-7B              & 6.12                 & 0.19              & 4.88              & 46.89             & 0.34              & 26.86             \\ 
    Deepseek-7B-SFT & \textbf{100.0}\textcolor{green}{[+93.88]}    & \textbf{0.50}\textcolor{green}{[+0.31]}   & 49.65\textcolor{green}{[+44.77]}   & 53.07\textcolor{green}{[+6.18]}    & \textbf{0.42}\textcolor{green}{[+0.08]}   & 34.91\textcolor{green}{[+8.05]}    \\  \midrule
    Mistral-7B               & 5.63                 & 0.11              & 2.58              & 35.34             & 0.25              & 24.19             \\ 
    Mistral-7B-SFT  & 96.65\textcolor{green}{[+91.02]}      & 0.47\textcolor{green}{[+0.36]}      & 49.14\textcolor{green}{[+46.56]}   & 47.17\textcolor{green}{[+11.83]}   & 0.37\textcolor{green}{[+0.12]}     & 29.01\textcolor{green}{[+4.82]}    \\  \midrule
    Qwen-2.5-72B             & 5.33                 & 0.27              & 3.26              & 40.00                & 0.28              & 29.95             \\ 
    Qwen-2.5-7B              & 12.14                & 0.24              & 5.21              & 43.50              & 0.25              & 27.85             \\  
    Qwen-2.5-7B-SFT & \textbf{100.0}\textcolor{green}{[+87.86]}    & 0.49\textcolor{green}{[+0.25]}     & \textbf{49.94}\textcolor{green}{[+44.73]} & 50.00\textcolor{green}{[+6.5]}     & 0.40\textcolor{green}{[+0.15]}     & 32.29\textcolor{green}{[+4.44]}    \\  \bottomrule
\end{tabular}

\end{table}

\subsection{Main Results}

We evaluated Intention Understanding, Intention Prediction, and Intention Execution on the ID test set. \textcolor{black}{All reported improvements compare fine-tuned models with their non-fine-tuned counterparts under the same agent framework, rather than with a separate baseline agent.}

\subsubsection{Evaluation of Intention Understanding}
Table~\ref{tab:Understander_Evaluation} presents the performance across the different models on the ID test set. \textcolor{black}{The results show that closed-source and non-fine-tuned models have limited performance in this setting. In contrast, the fine-tuned models improve intent segmentation and descriptive capabilities.} Specifically, Qwen-2.5-7B-SFT \textcolor{black}{improves} its grouping accuracy Acc-G from 12.14\% to 100\% and its \textcolor{black}{semantic accuracy} Acc-S from 0.24 to 0.49. \textcolor{black}{A similar performance gain is observed with Llama-3.1-8B, suggesting the usefulness of fine-tuning.}
\textcolor{black}{For intention understanding, both non-fine-tuned Qwen-2.5 models obtain very low Acc-G scores (5.33 for 72B and 12.14 for 7B), indicating that neither model can reliably perform intention segmentation without task-specific fine-tuning.} Although the post-SFT model achieves 100\% accuracy in intention segmentation (Acc-G), its Acc-S remains 0.49, \textcolor{black}{showing that correct boundary detection does not guarantee fine-grained semantic description. To examine this gap, we analyze 100 correctly segmented trajectories in Appendix~\ref{appendix:accs_error_analysis}. Most low-Acc-S cases are caused by over-generalization or missing details, suggesting a granularity mismatch between concise predictions and more specific reference annotations rather than a complete misunderstanding of user intentions.}

Furthermore, we conducted an ablation study to empirically examine whether action description quality limits downstream intention understanding. Specifically, we first generated action descriptions using three representative VLMs, GPT-4o, Gemini-1.5-pro, and Qwen-VL-Max, and measured their semantic alignment with the reference action descriptions using Acc-S. We then fed each set of generated descriptions into the same Qwen-2.5-7B-SFT$\dagger$ intention-understanding model. \textcolor{black}{This design isolates the effect of the description function while keeping the segmentation and intention-understanding model fixed. As shown in Table~\ref{tab:Understander_Ablation}, different VLMs lead to different downstream performance. However, the best generated descriptions, produced by Qwen-VL-Max, achieve an action-description Acc-S of 0.64 and lead to 92.91\% Acc-G and 0.45 Acc-S in intention understanding, which is close to the reference-description setting with 100.0\% Acc-G and 0.49 Acc-S.  These results suggest that, under our current setting, VLM-based action description is not the dominant bottleneck for segmentation. Therefore, we treat action description as an evaluated intermediate module rather than a separate benchmark task.
}

\begin{table}[htbp]
\centering
    \begin{minipage}[t]{0.48\textwidth}
    \centering
    \caption{Ablation experiment of Intention Understanding.}
    \label{tab:Understander_Ablation}
        \begin{tabular}{lc|ll}
        \toprule
        \multirow{2}{*}{Method} & Action Description & \multicolumn{2}{|l}{Understanding} \\ \cmidrule(r){2-2} \cmidrule(r){3-4}  
                                & Acc-S              & Acc-G           & Acc-S           \\ \midrule
        \makecell[l]{\textcolor{black}{Reference}\\\textcolor{black}{\hspace{0.6em}descriptions}}             & -                  & 100.0           & 0.49            \\ 
        GPT-4o                  & 0.62               & 92.17           & 0.38            \\
        Gemini-1.5-pro          & 0.54               & 89.03           & 0.35            \\
        Qwen-vl-max             & 0.64               & 92.91           & 0.45            \\ \bottomrule
        \end{tabular}
    \end{minipage}
\hfill
    \begin{minipage}[t]{0.48\textwidth}
    \centering
    \caption{Ablation experiment of Intention Prediction.}
    \label{tab:Ablation_Prediction}
        \begin{tabular}{llll}
        \toprule
        Methods                         & Acc-C      & Acc-S & BLEU-4 \\ \midrule
        \multicolumn{4}{l}{w/ understanding} \\ 
        \hspace{1em}GPT-4o                & 34.50   & 0.32 & 16.74  \\
        \hspace{1em}Qwen-2.5-7B-SFT       & 50.00   & 0.40 & 32.29  \\ \midrule
        \multicolumn{4}{l}{w/o understanding} \\ 
        \hspace{1em}GPT-4o                & 12.00   & 0.19    & 8.01   \\
        \hspace{1em}Qwen-2.5-7B-SFT       & 8.50    & 0.17    & 12.37   \\
        \bottomrule
        \end{tabular}
    \end{minipage}
\end{table}

\subsubsection{Evaluation of Intention Prediction}
The right part of Table~\ref{tab:Understander_Evaluation} presents the performance of different models in predicting intentions. Among closed-source models, Qwen-max achieves the highest category accuracy (48.5\%) and strong semantic consistency, while Gemini-1.5-pro attains the best semantic alignment with 0.36 Acc-S score. However, their overall performance still lags behind fine-tuned models. For open-source models, SFT also brings performance gains across all models. Specifically, Llama-3.1-8B-SFT achieves the best category accuracy (54.2\%) and BLEU-4 (34.95), marking an absolute improvement of +16.4 and +10.08 over the base model. These results indicate that lightweight fine-tuning on Act2Intention Bench effectively enhances the models’ ability to infer latent user intentions, even outperforming much larger proprietary models.

\textcolor{black}{For intention prediction, the lower Acc-C of the non-fine-tuned Qwen-2.5-72B suggests a mismatch between general-purpose generation and the benchmark-specific category taxonomy: without fine-tuning, larger models may produce more open-ended or fine-grained category descriptions that are semantically plausible but less aligned with the predefined category labels used for exact category evaluation.}

To further investigate the contribution of intent understanding to predictive performance, we conducted another ablation experiment comparing models with and without understanding modules. Without intention understanding, the model directly predicts the next intention from the raw action description sequence. We also reorganized the training data to fine-tune Qwen-2.5-7B for this setting. As shown in Table~\ref{tab:Ablation_Prediction}, \textcolor{black}{both GPT-4o and Qwen-2.5-7B-SFT show a clear performance decrease when the understanding module is removed. This decrease mainly stems from two factors.} First, predicting directly from low-level actions prevents the model from focusing on high-level intention semantics, making it harder to capture abstract user goals. Second, raw action sequences are very long (up to 12K tokens), adding noise and computational burden. \textcolor{black}{These results suggest that decoupling understanding (L2) and prediction (L3) is useful for the overall framework.}

\subsubsection{Comparison between Intention Understanding and Prediction} 

As shown in Table~\ref{tab:Understander_Evaluation}, when comparing the semantic metrics Acc-S and BLEU-4 between intention understanding and intention prediction, we observe a clear trend. For models without fine-tuning, intention understanding mostly achieves lower Acc-S and BLEU-4 scores than prediction. After supervised fine-tuning (SFT), however, understanding outperforms prediction. We attribute this to the difficulty of segmentation–summarization tasks in intention understanding for LLMs. Because LLMs have not been exposed to similar data during pre-training. As a result, the models struggle to segment user actions, further leading to inaccurate semantic intent generation. \textcolor{black}{Compared with intention prediction, intention understanding appears more learnable in our setting. Fine-tuning on Act2Intention Bench improves LLM performance on intention understanding.}

\subsubsection{Evaluation of Intention Execution}
As shown in Table~\ref{tab:Intention_Execution}, \textcolor{black}{we evaluate Experience-Guided Intention Execution across multiple GUI Agents.} Since Act2Intention Bench consists of OOD data for UI-TARS, both the UI-TARS-2B-SFT and UI-TARS-7B-SFT models exhibit relatively low step accuracy. However, by incorporating relevant action trajectories into the prompts, their Step Success Rate (SSR) improves by 9.9 and 3.2, respectively. Notably, the 2B model with experience guidance shows a larger improvement, achieving an SSR comparable to the 7B model without guidance. This may be because the 2B model has weaker subtask decomposition capabilities, and the guidance information helps strengthen this aspect. Beyond UI-TARS, both GUI-R1-3B and CogAgent-9B \textcolor{black}{show gains} of +6.8 and +7.7 SSR, respectively. However, compared with UI-TARS-2B-SFT and CogAgent-9B, the 2B model exhibits better overall performance and larger performance gains. Compared to CogAgent, UI-TARS benefits from training on a larger amount of data and the incorporation of long-term memory and reflective adjustment mechanisms. \textcolor{black}{Overall, the improvements across different agents suggest that the proposed method can be applied to multiple GUI-agent backbones.}

\subsubsection{Evaluation of End-to-End} 

Finally, we evaluate the complete pipeline of Act2Intention, where the agent first understands the historical GUI trajectory, predicts the next user intention, and then executes the predicted intention with a mobile GUI executor. \textcolor{black}{We report two metrics: Acc-S for the semantic accuracy of predicted intentions, and SR for the final task success rate after online execution in the emulator. Unlike SSR, SR measures whether the intended task is completed after interaction and does not require the executed action path to exactly match the reference trajectory, thereby allowing functionally equivalent GUI paths.} \textcolor{black}{Rather than treating the end-to-end result merely as a final system score, this experiment is intended to reveal where the current proactive mobile agent pipeline remains challenging.}

As shown in Table~\ref{tab:End-to-End}, the best end-to-end SR reaches 22.7 when using Llama-3.1-8B-SFT for understanding/prediction and UI-TARS-7B-SFT as the executor.\textcolor{black}{This result is lower than the oracle-intent execution SSR in Table~\ref{tab:Intention_Execution}, suggesting that fully autonomous task completion is more difficult than reactive execution with a given instruction. To analyze this gap, we further introduce a ground-truth intention setting, where the predicted intention is replaced with the annotated one. With ground-truth intentions, the SR improves from 22.7 to 47.6 for UI-TARS-7B-SFT. This indicates that predicted intentions are often not accurate enough. Meanwhile, the SR remains limited, suggesting that long-horizon GUI execution itself is another bottleneck. These results highlight a key challenge for proactive mobile agents: a complete system needs to jointly address reliable intention prediction and robust execution under long action chains, diverse app contexts, and imperfect intermediate representations.}

\textcolor{black}{
To further examine a practical mitigation strategy, we evaluate confidence-based fallback using UI-TARS-7B-SFT as the executor. 
\textcolor{black}{Before pushing the predicted intention to the user, the agent calls a confidence evaluator to output a confidence score, verifying whether the prediction is likely to satisfy the user's actual need and executable (See Appendix~\ref{prompt:ConfidenceFallback} for details). The decision rule is explicit: a predicted intention is pushed only when its confidence score is no lower than the default threshold $\theta=0.60$; otherwise, it is discarded by fallback.} As shown in Table~\ref{tab:End-to-End-Fallback}, this mechanism reduces unreliable suggestions and improves the success rate among pushed cases. 
}

\textcolor{black}{
As a future improvement, the discarded low-confidence prediction and the annotated next user intention can be naturally collected as a preference pair, where the former serves as a rejected response, and the latter serves as a preferred response for DPO-style training. In practical deployment, the system can further provide a lightweight confirmation interface, where users accept, reject, or revise the predicted intention before execution. Such user feedback can also be collected as preference data to further align the prediction model with user needs.
}

\vspace{-5mm}

\begin{table}[htbp]
\centering
\begin{minipage}[t]{0.43\textwidth}
    \centering
    \caption{Evaluation results of \textbf{Intention Execution}.}
    \label{tab:Intention_Execution}
    \small
    \setlength{\tabcolsep}{2pt}
    \begin{tabular}{lcll}
    \toprule
    Methods                         & Guidance      & SSR & Time \\ \midrule
    \multirow{2}{*}{UI-TARS-2B-SFT} & \ding{55}        & 47.0        &  3.55            \\
                                    & \ding{51}        & 56.9\textcolor{green}{[+9.9]}    &  3.76    \\ \midrule
    \multirow{2}{*}{UI-TARS-7B-SFT} & \ding{55}        & 59.6        &  6.18            \\
                                    & \ding{51}        & \textbf{62.8}\textcolor{green}{[+3.2]}    &  6.56     \\ \midrule
    \multirow{2}{*}{GUI-R1-3B}     & \ding{55}         & 42.3        &  4.00            \\
                                    & \ding{51}        & 49.1\textcolor{green}{[+6.8]}        &  4.08            \\ \midrule
    \multirow{2}{*}{CogAgent-9B} & \ding{55}           & 48.3        &   6.46           \\
                                    & \ding{51}        & 56.0\textcolor{green}{[+7.7]}        &   6.90           \\  \bottomrule
    \end{tabular}
\end{minipage}
\hfill
\begin{minipage}[t]{0.55\textwidth}
    \centering
    \caption{\textcolor{black}{End-to-end evaluation with predicted and g.t. intentions.}}
    \label{tab:End-to-End}
    \small
    \setlength{\tabcolsep}{2pt}
    \begin{tabular}{lllll}
    \toprule
    Understanding                              & Prediction                                     & Execution      & Acc-S                     & SR   \\ \midrule  
    \multirow{3}{*}{Qwen-2.5-7B-SFT$\dagger$}  & \multirow{3}{*}{Qwen-2.5-7B-SFT$\ddagger$}     & UI-TARS-7B-SFT & \multirow{3}{*}{0.32}     & 20.9 \\
                                               &                                                & GUI-R1-3B      &                           & 15.5 \\
                                               &                                                & CogAgent-9B    &                           & 13.2 \\ \midrule  
    \multirow{3}{*}{Llama-3.1-8B-SFT$\dagger$} & \multirow{3}{*}{Llama-3.1-8B-SFT$\ddagger$}    & UI-TARS-7B-SFT & \multirow{3}{*}{0.31}     & 22.7 \\
                                               &                                                & GUI-R1-3B      &                           & 15.7 \\
                                               &                                                & CogAgent-9B    &                           & 12.5 \\ \midrule  
    \multirow{3}{*}{\textcolor{black}{-}}        & \multirow{3}{*}{\textcolor{black}{Ground Truth}} & \textcolor{black}{UI-TARS-7B-SFT} & \multirow{3}{*}{\textcolor{black}{1.00}}  & \textcolor{black}{47.6} \\
                                               &                                                & \textcolor{black}{GUI-R1-3B}      &                                         & \textcolor{black}{39.2} \\
                                               &                                                & \textcolor{black}{CogAgent-9B}    &                                         & \textcolor{black}{33.5} \\
    \bottomrule
    \end{tabular}
\end{minipage}
\vspace{-2mm}
\end{table}

\vspace{-5mm}

\section{LIMITATIONS AND FUTURE DIRECTIONS} 

This work primarily contributes a new benchmark and a framework for proactive mobile agents, while most constituent modules are built upon existing techniques, such as supervised fine-tuning LLMs. \textcolor{black}{Here, our goal is to examine the feasibility of the \textit{Act2Intention} framework and provide an evaluation setting for proactive agents.} However, there is still room for improvement in both intent prediction and execution. In intention prediction, future work can use long-term memory (e.g., user intentions at the same time in previous days or weeks) and leverage more contextual signals such as user emotion, weather, and location. For intention execution, future directions include integrating world models \cite{ding2025understanding} to enable the agent to learn adaptive and long-horizon decision-making strategies. 

\textcolor{black}{
Another limitation is that Act2Intention focuses on anticipatory intention prediction rather than opportune intervention timing. In designing the current benchmark, we did not include labels or tasks for deciding when an agent should interrupt or engage the user. Accordingly, our prototype surfaces predicted intentions through predefined system events, such as device wake-up, and requires user confirmation before execution. Future extensions could add an \textit{opportune-moment prediction} task that jointly evaluates \textbf{what} intention should be suggested and \textbf{when} the suggestion should be presented, using contextual signals such as current activity, interaction state, notification history, and user feedback.
}

\textcolor{black}{
A further limitation concerns user-centric evaluation. Although Act2Intention provides quantitative metrics for intention understanding, prediction, and execution, we have not yet evaluated whether the agent's proactive suggestions are perceived as useful, timely, and non-intrusive by real users. Prior HCI studies on human-AI interaction and proactive assistants show that user acceptance depends on suggestion relevance, timing, and user control~\cite{amershi2019guidelines,oh2024better,berube2024proactive}. In future work, we plan to conduct in-situ user studies to evaluate how users accept, reject, or correct proactive suggestions. We will also compare the current top-1 suggestion design with a potential top-$K$ variant to examine whether multiple candidate intentions improve usefulness or introduce additional cognitive burden.
}

\textcolor{black}{
Act2Intention also has limitations in dataset coverage and generalization. The current benchmark covers 52 apps and common mobile usage patterns from 90 anonymous participants. Due to privacy agreements, we do not have access to demographic information, such as age, gender, region, or technical background. Therefore, the collected behaviors may reflect specific user groups or usage habits. Although synthetic trajectories increase the scale and diversity of the benchmark, they cannot fully replace real behavior from diverse populations. They may also contain idealized or stereotypical patterns introduced by LLM generation. In addition, the agent may struggle with unseen apps, new UI layouts, or cross-app long-horizon intentions. Future work should collect data from more diverse users, evaluate demographic and behavioral bias, and study cross-app generalization and continual adaptation.
}

\textcolor{black}{
Finally, real-world deployment raises both privacy and efficiency challenges. Although the data used in this work were anonymized and desensitized during collection, practical deployment may involve sensitive screenshots, actions, and personal usage histories. Current large-model-based agents may also introduce latency and energy overhead when running UI parsing, intention prediction, and GUI execution on mobile devices. Our experiments are conducted in emulator-based settings, and real-device performance may be slower. Future deployments should explore stronger privacy protection, such as secure storage, data anonymization, differential privacy, federated learning, and edge collaboration. They should also reduce computation cost through model compression, quantization, smaller on-device models, and lightweight screenshot redaction before data transmission. These directions are important for moving Act2Intention from benchmark evaluation to practical proactive mobile assistance.
}

\section{CONCLUSION}
\textcolor{black}{In this work, we study proactive mobile agents centered on ``Understanding $\rightarrow$ Predicting $\rightarrow$ Executing'' user intentions. We constructed Act2Intention Bench through collection and validated automated generation. It comprises 72,511 intentions and over 700,000 actions across 52 apps, providing a benchmark for evaluating proactive agents via continuous ``intention-actions'' trajectories. Using Act2Intention Bench, we developed the Act2Intention Agent, which achieves competitive performance in intent understanding and prediction, suggesting the benchmark's value for training and evaluating proactive agents. We hope this work offers a useful step toward more capable proactive mobile assistants.}

\begin{acks}
This work was supported by the National Key R\&D Program of China (No. 2024YFB4505502) and the National Natural Science Foundation of China (No. 62441229, No. U24B20180).

\end{acks}

\bibliographystyle{ACM-Reference-Format}
\bibliography{sample-base}

\newpage
\appendix

\section{ETHICS STATEMENT} \label{appendix:Ethics}

Our process of constructing the dataset conforms to the code of ethics. Participants in data collection were explicitly informed about the study’s purpose of intention inference, and were made aware of potential privacy risks. The released dataset has been processed with safety filters to ensure that there is no harmful content or private information in our dataset.

\section{STATEMENT ON THE USE OF GENERATIVE AI} 

This paper involved the use of generative artificial intelligence tools for language improvement only. Specifically, we used the GPT-5, DeepSeek, and Grammarly to correct grammatical errors, enhance sentence clarity, and improve logical coherence between paragraphs.

\section{ACT2INTENTION FRAMEWORK DETAILS}

\subsection{Data Collection} \label{appendix:data_collection}

\begin{figure}[htb]
  \centering
  \includegraphics[width=0.7\linewidth]{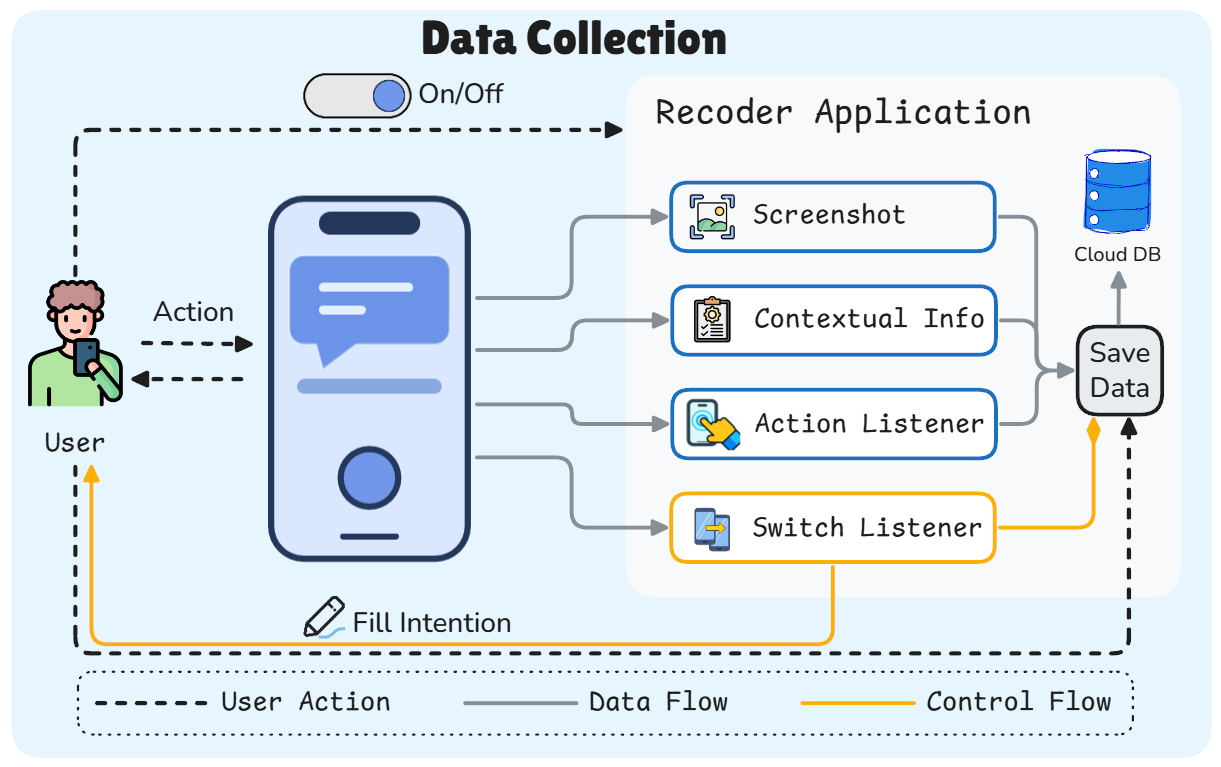}
  \caption{\textcolor{black}{Overview of the data collection process.}}
  \label{fig:data_collection}
\vspace{-4mm}
\end{figure}

\textcolor{black}{Figure~\ref{fig:data_collection} illustrates the data collection procedure used by the smartphone manufacturer. Specifically, after the participant activated the recording function, the record application monitored foreground app-switching events during normal phone usage. Whenever an app-switching event was detected, the recorder prompted the participant with a dialog box to briefly describe their intention for using the current application. Meanwhile, the recorder collected contextual information, screenshots, and a sequence of user actions through Android API calls, including action types, click coordinates, input text, swipe directions, and other operation metadata. The recorder continued collecting these records until the next app-switching event was detected, thereby forming one user-stated intention--action unit. This process was repeated until the participant manually stopped the recording session, resulting in a continuous intention--action trajectory as one data instance. Before release, each user-provided intention is screened to remove private or sensitive information and then normalized into a concise, task-oriented description while preserving its original meaning. All collected data were desensitized, screened for safety, and securely uploaded to the cloud server by the smartphone manufacturer before being shared with our research team.}

\subsection{Data Generation} \label{appendix:data_generation}

\begin{AIbox}{User Persona Generation}
<Role>You are a user data analyst</Role>
<Task>Your task is to generate a comprehensive and personalized user profile description based on the user's historical mobile phone usage intent trajectory.</Task>
<Rule>  
0. The input format is: \texttt{List[\{"Time": "string", "APP": "string", "Intention": "string", "Event": "string"\}]}, where "Event" represents the category of each user intent.  
1. First, you should analyze the intentions and their corresponding event categories to determine which events are semantically similar.  
2. Based on a clear understanding of the event semantics, merge high-frequency sub-items according to their semantic similarity. The merging process must adhere to the following requirements:  
   (1) Carefully consider whether the semantics of the sub-items to be merged are truly identical and whether there is a better way to merge them.  
   (2) Each merge will result in slightly more generalized semantics. Pay attention to the number of merges:  
      - Merging too much will make the semantics overly broad and vague, applicable to most users, and failing to reflect the individual's unique traits.  
      - Merging too little will make the behavior patterns overly fragmented and detailed, obscuring the user's core behavior patterns.  
   (3) Before each merge, you may refer to the timestamps of the sub-items in the historical behavior sequence to better inform your judgment. Additionally, if a behavior pattern exhibits clear temporal characteristics, state them directly.  
3. Finally, based on the historical behavior sequence and your merged behavior patterns, provide a thorough analytical summary of the user's profile. The summary should be comprehensive while highlighting the user's personalized traits.  
4. Maintain an objective description and avoid excessive speculation. All inferences should be grounded in the user's actual historical intents and behavior patterns. Rather than describing highly uncertain imagined content, focus only on what can be confidently deduced from the user's historical intent sequence and behavior patterns. Minimize aesthetic descriptions as much as possible.  
5. All the Intentions mentioned above are collected from users' mobile phone usage.
6. Strictly output in JSON format: \texttt{\{"Persona": str\}}, and DO NOT output anything other than JSON.
</Rule>
\end{AIbox}

\begin{AIbox}{Intention Trajectory Generation}
<Role>You need to act as a real user.</Role>  
<Task>Your task is to generate realistic app usage intention trajectories based on the user profile and scenario I provide.</Task>  
<Rule>  
1. Each intention must include a specific Time, a concrete App name, and a realistic Intention.  
2. Intentions should follow a logical flow and connect coherently.  
3. Ensure each intention contains only one brief and specific action within an app; avoid combining multiple actions in a single description.  
4. Objectively describe smartphone usage intentions without additional content [e.g., reasoning, interpretations, or subjective opinions].  
5. Ensure all mentioned apps and their corresponding operations are realistic and feasible.  
6. Strictly output in JSON format: \texttt{List[\{"Time": "string", "APP": "string", "Intention": "string"\}]},  and DO NOT output anything other than JSON.
</Rule>
<Persona>
The user is a 29-year-old female who lives in urban and works in a technology-driven field. She tends to prefer online shopping over in-store experiences, favoring convenience, personalized recommendations, and high-quality or sustainable products, often browsing fashion, home décor, and wellness items. Her lifestyle is structured yet flexible; she starts her day with light exercise such as yoga or jogging, enjoys preparing healthy meals at home, and values work-life balance, often dedicating evenings to reading, streaming, or socializing with close friends. She has a range of hobbies, including photography, traveling to culturally rich destinations, cooking new recipes, and engaging in creative projects like painting or digital design. Her primary electronic devices include a iphone 15 promax, a ipad for media, a laptop for work, and wireless earbuds. She values personal growth, sustainability, curiosity, and authenticity, prefers practical problem-solving, and holds progressive social and environmental views, with no particular religious affiliation. She owns a small dog  and she is attentive to her health, maintaining regular fitness routines and mindful nutrition habits. Her daily app usage includes social media platforms such as Instagram and TikTok, productivity apps like Notion and Google Calendar, wellness apps including Calm and MyFitnessPal, streaming services like Spotify and Netflix, and shopping apps such as Amazon and Etsy, reflecting a digitally integrated, health-conscious, and lifestyle-oriented persona.
</Persona>
\end{AIbox}

\begin{AIbox}{Trajectory Validation} \label{appendix:data_check}
<Role>
You are a rigorous evaluator of mobile UI action trajectories. Given an intention, an ordered action trajectory (each step include action\_type, app\_name, x, y, description, image filename), and the final-frame screenshot, decide whether the trajectory achieves the intention. Return exactly one JSON object and nothing else.
</Role>  
<Task>
Assess whether the provided action trajectory successfully accomplishes the given intention.
</Task>
<Rule> 
Base your judgment only on the intention, the step metadata (including descriptions), and the final screenshot.
    - Be strict: if critical steps are missing or the final state is uncertain, mark the trajectory as not successful.
    - Prefer explicit step descriptions over implicit coordinate clicks; x/y without meaningful descriptions are weak evidence.
    - Handle minor schema typos (e.g., 'decsription' -> 'description') logically.
    - Consider completeness (all necessary sub-goals covered), correctness (actions align with the intention), and final state plausibility.
    - Do not include any extra text besides the JSON.
Scoring rubric (score 0-100):
    - 90-100: Goal clearly achieved; steps are coherent and sufficient; final state strongly consistent with the intention.
    - 60-89: Largely correct with minor gaps/uncertainties; likely achieved but not fully evidenced.
    - 1-59: Partially or poorly executed; important steps missing; unlikely achieved.
    - 0: No meaningful progress or clearly unrelated to the intention.
    Confidence (0-1): Reflect your certainty from evidence strength (higher when steps/descriptions strongly support success).
    Output JSON schema (return only this object):
    {
        "success": true|false,
        "reason": "brief one-sentence reason",
        "score": number (0-100),
        "confidence": number (0-1)
    }
</Rule> 
<Instruction></Instruction>
<Trajectory></Trajectory>

\end{AIbox}

\subsection{\textcolor{black}{Human Realism Assessment Details}}
\label{appendix:human_realism}

\textcolor{black}{
We recruited 10 annotators to evaluate the realism of sampled ``intention-actions'' trajectories. The annotators were graduate students or senior undergraduate students. All annotators worked independently and were blind to the source subset of each sample, i.e., they did not know whether a sample came from RR, RG, or GG. For the assessment, we randomly sampled 30 intention segments from each subset, resulting in 90 samples in total. Specifically, since RR and RG are constructed from the same 90 behavior-derived personas, we first randomly sampled 10 shared personas from these 90 personas. Similarly, we sampled 10 personas from the generated-persona subset. For each selected persona, we randomly sampled 3 ``intention-actions'' trajectories from RR, RG, and GG. Each sample contained the app name, intention trajectories, action description sequence, and the associated behavior-derived persona. }

\textcolor{black}{
Annotators rated each sample along three dimensions using a 5-point Likert scale. \textit{Intention realism} measures whether the intention resembles a plausible real-world mobile usage goal. \textit{Action achievability} measures whether the action trajectory can reasonably accomplish the given intention. \textit{Persona consistency} measures whether the intention is consistent with the associated behavior-derived persona. \textcolor{black}{Given this limited scale, the results provide preliminary evidence of plausibility rather than a comprehensive validation of realism, diversity, or bias.}
}

\subsection{Act2Intention Agent} \label{appendix:Act2Intention_Agent}

\begin{AIbox}{Action Description} \label{prompt:ActionDescriber}
<Role>You are a professional GUI action Describer</Role>
<Task>You are given an action and the split-screenshot image (Before-action left, After-action right) on the mobile phone. Then, you need to describe the action.</Task>
<Rule>
- For click actions, a high-contrast red marker (white-bordered circle) shows the precise click location, with a green square surrounding it and a ’C’ label at the top-right corner of the square indicating the click.
- Strictly output in JSON format: \texttt{\{"Action\_Description": str\}}
- The "Action\_Description" field in this exact format: \texttt{"[On/In] [PackageName], [Action Details], to [Purpose]"}
- The "Action\_Description" field needs to keep within 20 English words
- DO NOT output anything other than JSON
</Rule>
\end{AIbox}

\begin{AIbox}{Intent Understanding} \label{prompt:IntentParser}
<Role>You are a mobile action descriptions analysis expert, responsible for identifying user intentions on mobile devices. </Role>
<Task> You are tasked with grouping action description sequences and identifying the corresponding intention for each group. You need output in JSON format.</Task>
Please process user input according to the following rules:
<Rule>
1. Intention Segmentation Analysis:
    - Identify the boundaries between different intentions in a continuous action description sequence.
    - Determine intention switches based on user intent, application scenarios, and temporal continuity.
    - Each independent intention must contain at least ONE action description.
2. Intention Description Standards
    - Use a verb-object structure (verb + target object) (e.g., "Modify system settings")
    - Include core verbs and application scenarios
    - Avoid using the exact wording from the action description steps
3. Formatting Requirements:
    - Strictly output in JSON format: \texttt{\{string: List[int]\}}.
    - The output is a dictionary with Intention Descriptions as keys and a list of Action Description INDICES as values.
    - Each Intention Description is a list of action description indices.
    - Index starts at 1.
    - DO NOT output anything other than JSON.
</Rule>
<Format> Strictly output in JSON format: \texttt{\{"Intention Description": [steps]\}}! </Format>
\end{AIbox}

\begin{table}
    \centering
    \caption{Action space.}
    \begin{tabular}{ll}
        \toprule
         Action             & Description     \\ \midrule
         CLICK[x, y]        & Click on the position with coordinates [x, y]     \\
         LONG\_CLICK[x, y]  & Long-click at the position with coord. [x, y]     \\
         TYPE[text]         & Input the content of ``text'' in the input box    \\
         SWIPE[direction]   & Swipe in the specified direction                   \\
         PRESS[KEY]         & Press a key in ``HOME'',``ENTER'',``BACK''\\ \bottomrule
    \end{tabular}
    \label{tab:actions_Description}
\end{table}

\begin{algorithm}
\caption{Intent Understanding Process}
\label{alg:IntentParser_appendix}
\begin{algorithmic}[1]
    \Require Action Description Memory $\mathbf{M}_D = \{  \mathcal{D}_1^*,\ldots,\mathcal{D}_{t-1},\mathcal{D}_t \}$
    \Ensure Intention Memory $\mathbf{M}_I$
    \State $\mathbf{M}_I  \gets  \emptyset , N \gets 70 $
    \Loop
        \If{$\text{count}(\mathbf{M}_D.\text{unmarked}) \geq N$}
            \State $\tau \gets $ Retrieve the latest $N$ unmarked action descriptions $\{ \mathcal{D}_{t-N+1},\ldots,\mathcal{D}_t\}$ from $\mathbf{M}_D$
            \State $(\mathcal{I}_1, \tau_1),\dots,(\mathcal{I}_k, \tau_k) \gets A_U(\tau)$ Grouping and describing $k$ intentions  \Comment{Appendix~\ref{prompt:IntentParser}} 
            \For{$j \gets 1$ to $k-1$} \Comment{Mark the first $k-1$ groups of action descriptions}
                \ForAll{$\mathcal{D} \in \tau_j$}
                    \State $\mathbf{M}_D.\text{mark}(\mathcal{D})$
                \EndFor
            \EndFor
            
            \State $\mathbf{M}_I.\text{extend}( [ \mathcal{I}_1,\ldots,\mathcal{I}_{k-1}])$ 
        \EndIf
    \EndLoop
\end{algorithmic}
\end{algorithm}

\begin{AIbox}{Intent Prediction} \label{prompt:IntentPredictor}
<Role> You are a helpful assistant that provides proactive suggestions to the user. </Role>
<Task> Understand what the user is doing and predict their next intention based on historical intentions.</Task>
<Format> 
- Strictly respond in the following JSON format:
\texttt{"Event": "EVENT class to which intention belongs", "Behaviour": "Describe the predicted intention."\}}
- DO NOT output anything other than JSON.
</Format>
<Rules>
- Ensure the predicted intention is relevant to the historical intentions. 
- Focus on the user's current needs and predict helpful intentions.
- Consider the timing of Behaviour and the EVENT classes.
</Rules>
\end{AIbox}

\begin{algorithm}
\caption{IntentPredictor Process}
\label{alg:IntentPredictor_appendix}
\begin{algorithmic}[1]
    \Require{Intention Memory $\mathbf{M}_I = \{\mathcal{I}_1, \ldots, \mathcal{I}_t\}$, history length $m$}
    \Ensure{Predicted intent $\mathcal{I}'$}
    
    \State Fetch historical intentions: $\{\mathcal{I}_{t-m+1}, \ldots, \mathcal{I}_t\} \gets \mathbf{M}_I[-m:]$ 
    \State Predict latent intent: $\mathcal{I}' \gets A_P( \{\mathcal{I}_{t-m+1}, \ldots, \mathcal{I}_t\} )$ \Comment{Appendix~\ref{prompt:IntentPredictor}}
    
    \If{DeviceState == \textnormal{"ACTION\_SCREEN\_ON"}}
        \State Trigger proactive service notification with $\mathcal{I}'$
        \If{UserResponse == \textnormal{"Accept"}}
            \State Deploy $\mathcal{I}'$ through Executor
        \EndIf
    \EndIf
    \State \Return $\mathcal{I}'$ \Comment{Always return prediction}
\end{algorithmic}
\end{algorithm}

\begin{AIbox}{Confidence-based Fallback} \label{prompt:ConfidenceFallback}
\textcolor{black}{
<Role>You are a conservative evaluator for proactive mobile intention suggestions.</Role>
<Task>Given the user's historical intentions, user persona, current context, and the predicted intention, estimate whether the predicted intention is reliable enough to be pushed to the user.</Task>
<Input>
The input contains four parts:
1. \texttt{Persona}: a behavior-derived user profile.
2. \texttt{History}: recent historical intentions in the format \texttt{List[\{"Time": "string", "APP": "string", "Intention": "string", "Event": "string"\}]}.
3. \texttt{CurrentContext}: the current time, latest used app, and latest completed intention if available.
4. \texttt{PredictedIntention}: the proactive suggestion generated by the intention prediction model, in the format \texttt{\{"Event": "string", "Behaviour": "string"\}}.
</Input>
<Rule>
1. Evaluate whether the predicted intention is likely to satisfy the user's actual next need based on the persona, historical intentions, and current context.
2. Evaluate whether the predicted intention contains sufficient executable information, including the target app, operation, and expected outcome.
3. Be conservative. If the prediction is vague, over-general, weakly related to the user's history, or lacks key executable details, assign a low confidence score.
4. Do not assume information that is not supported by the input.
5. \textcolor{black}{The final decision should be \texttt{"push"} only when the prediction is both user-relevant and executable and its confidence score is no lower than the threshold $\theta=0.60$. Otherwise, the decision should be \texttt{"discard"}.}
6. Use the following confidence scale:
   - \textcolor{black}{0.80--1.00: highly relevant and executable; safe to push.}
   - \textcolor{black}{0.60--0.79: moderately relevant and executable; push under the default threshold $\theta=0.60$.}
   - 0.00--0.59: irrelevant, vague, or not executable; discard.
7. Strictly output in JSON format:
\texttt{\{"confidence": float, "decision": "push"|"discard", "reason": "brief reason"\}}.
8. DO NOT output anything other than JSON.
</Rule>
}
\end{AIbox}

\section{SUPPLEMENTARY EXPERIMENTS}

\subsection{Error Analysis of Semantic Accuracy under Correct Segmentation}
\label{appendix:accs_error_analysis}

\begin{figure}[htbp]
	\centering
	\begin{minipage}[c]{0.48\textwidth}
		\centering
		\includegraphics[width=\textwidth]{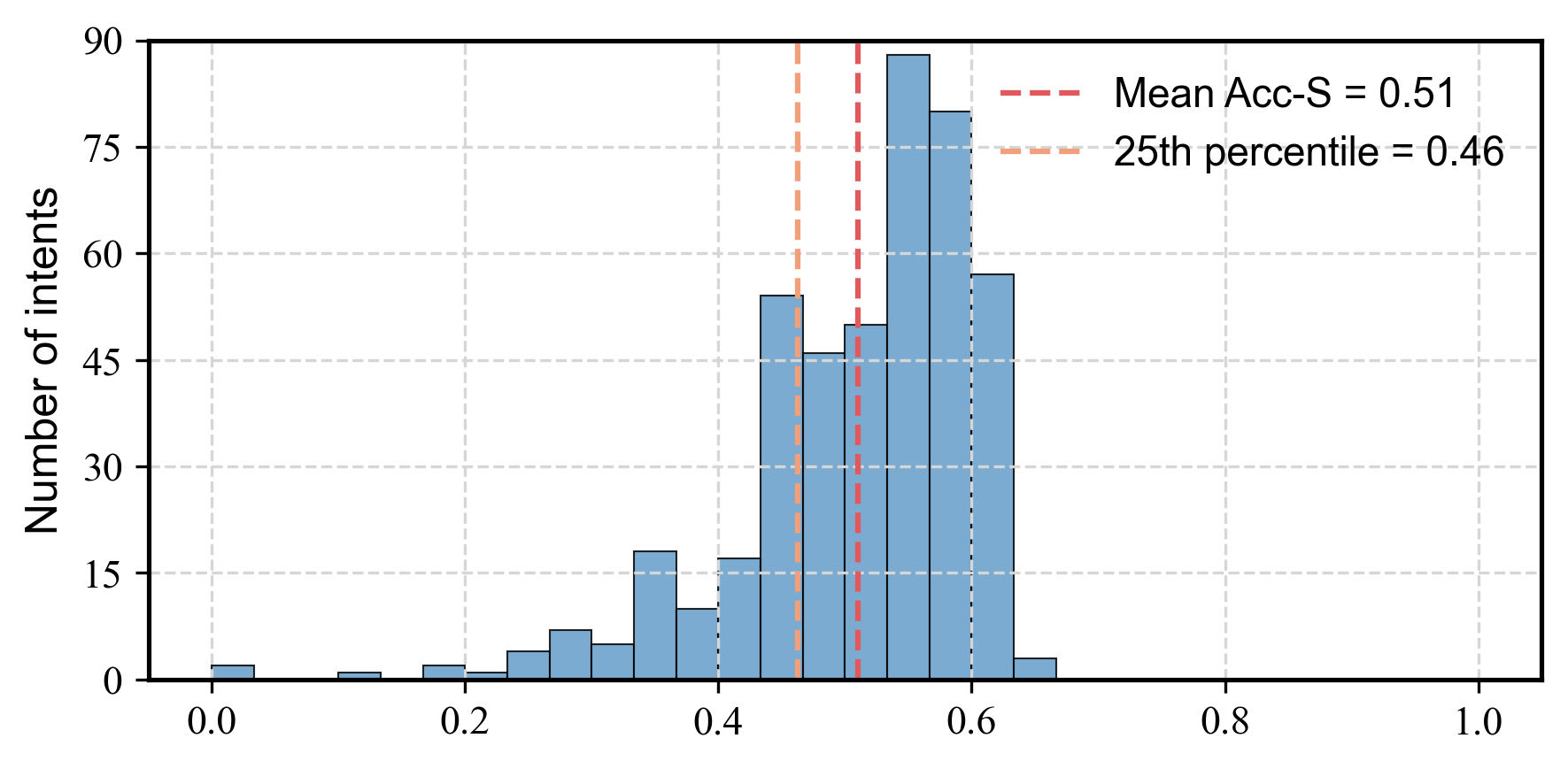}
		\subcaption{\textcolor{black}{Distribution of Acc-S.}}\label{fig:Distribution_AccS}
	\end{minipage} 
	\begin{minipage}[c]{0.48\textwidth}
		\centering
		\includegraphics[width=\textwidth]{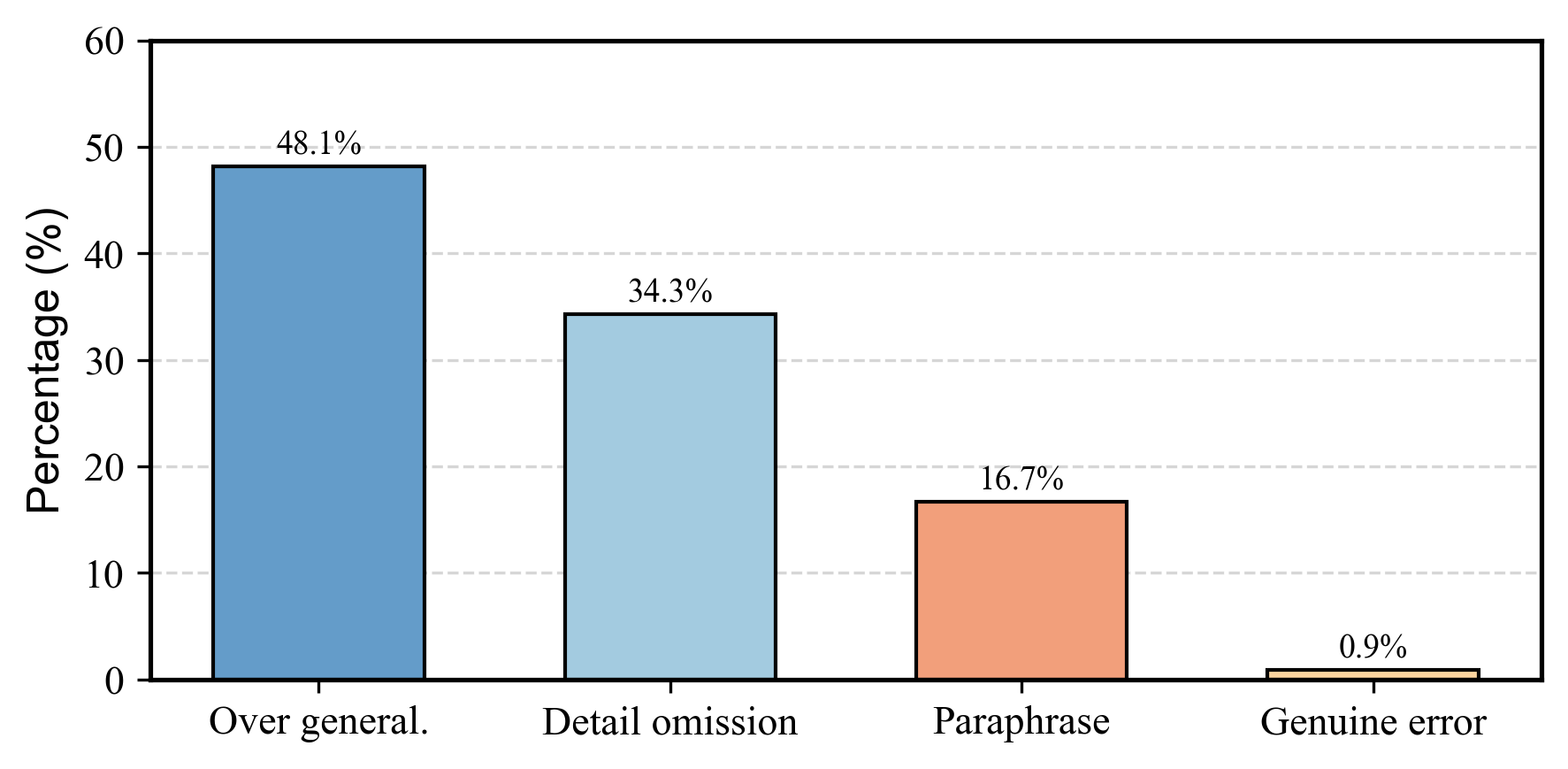}
		\subcaption{\textcolor{black}{Error types of Low Acc-S. }}\label{fig:proportions_types}
	\end{minipage}
    \caption{\textcolor{black}{Error analysis of semantic accuracy under correct segmentation.}}
  \label{fig:accs_error_analysis}
\end{figure}

\textcolor{black}{
To better understand the discrepancy between high grouping accuracy (Acc-G) and relatively low semantic accuracy (Acc-S), we conducted an additional analysis on 100 randomly sampled trajectories whose intention boundaries were correctly segmented. For each trajectory, we computed the semantic cosine similarity between the predicted intention and the reference intention. Figure~\ref{fig:accs_error_analysis} summarizes the results. Figure~\ref{fig:Distribution_AccS} shows the distribution of Acc-S over these correctly segmented cases. We then focused on the low-Acc-S subset, defined as samples below the 25th percentile of the Acc-S distribution, and manually categorized their semantic deviations into four types: \emph{over-generalization}, \emph{detail omission}, \emph{paraphrase / alternative wording}, and \emph{genuine semantic error}. Figure~\ref{fig:proportions_types} reports the proportion of each category. We observe that low-Acc-S cases are dominated by over-generalization and detail omission, indicating that many failures arise because the model captures the coarse intention correctly but omits app-specific, object-specific, or contextual details. This suggests that the gap between Acc-G and Acc-S reflects not only semantic weakness, but also a granularity mismatch between concise model outputs and more specific reference annotations.
}

\textcolor{black}{
Related intent-generation tasks also report task-dependent embedding-based semantic similarity ranges, e.g., 0.04--0.492 for zero-shot intent discovery and 0.61--0.862 for SBERT-based intent summarization \cite{Comi2023ZeroShotBERTAdapters,Yang2025FCMIR}.
}

\subsection{Confidence Statistics for Fallback}
\label{appendix:confidence_statistics}

\begin{figure}[htbp]
  \centering
  \includegraphics[width=0.55\linewidth]{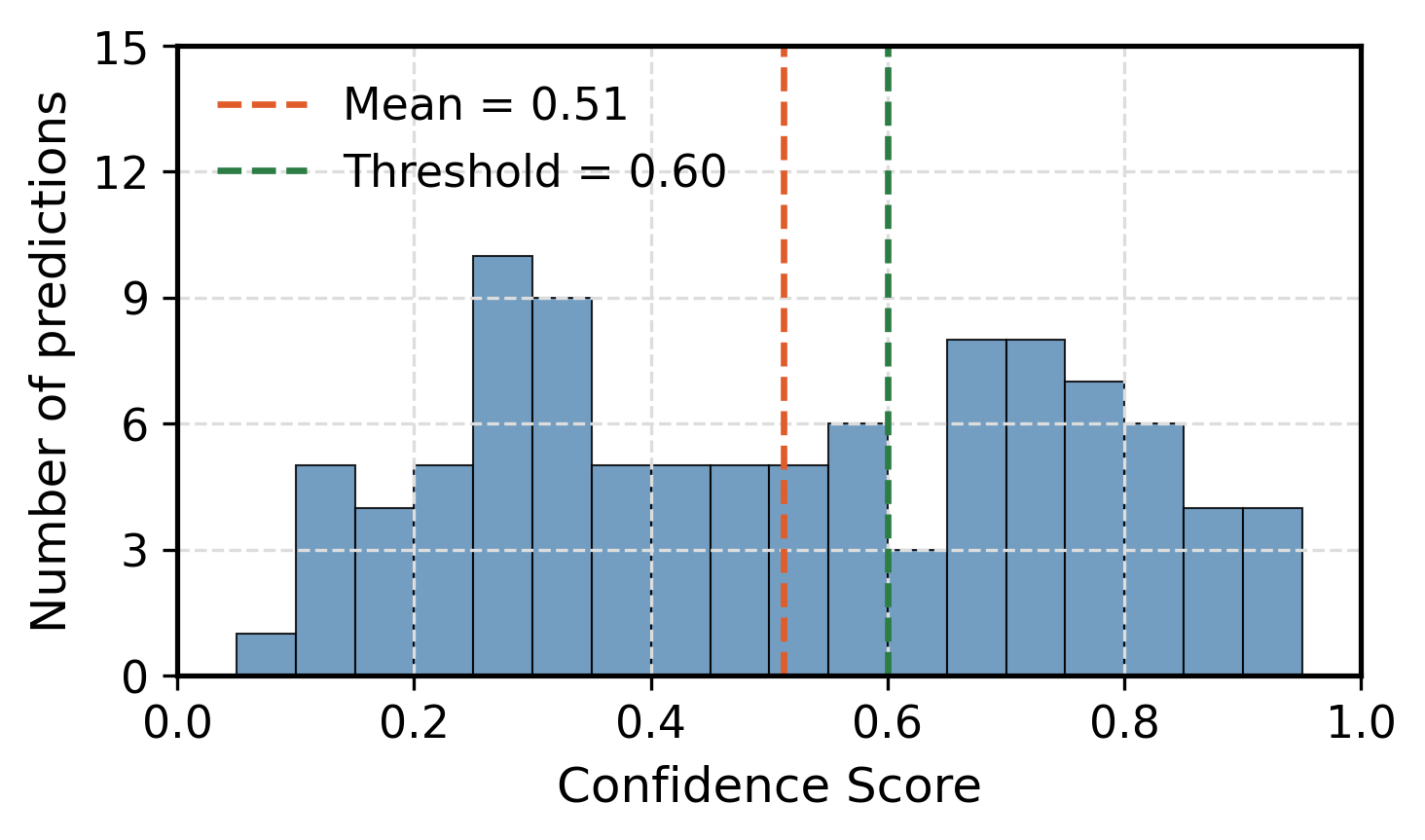}
  \caption{Distribution of confidence scores. The dashed green line denotes the threshold $\theta=0.60$.}
  \label{fig:confidence_distribution}
  \vspace{-2mm}
\end{figure}

\textcolor{black}{In this fallback experiment, we use Qwen-2.5-7B-SFT$\dagger$ for intention understanding, Qwen-2.5-7B-SFT$\ddagger$ for intention prediction, and UI-TARS-7B-SFT for executing the pushed intentions. Push Rate denotes the proportion of predicted intentions whose confidence score is above the threshold and are therefore pushed for execution, while the remaining predictions are withheld by fallback.} \textcolor{black}{In the table, Push Rate is reported as a percentage.} \textcolor{black}{Here, Acc-S$_{\text{push}}$ and SR$_{\text{push}}$ report the semantic accuracy and execution success rate computed only on the pushed samples.}
\textcolor{black}{Figure~\ref{fig:confidence_distribution} shows the distribution of confidence scores produced by the confidence evaluator in the end-to-end fallback experiment. The dashed green line denotes the threshold $\theta=0.60$.}

\begin{table}[htbp]
\centering
\caption{\textcolor{black}{Confidence-based fallback with Qwen-2.5-7B-SFT and UI-TARS-7B-SFT. The threshold $\theta=0.60$ is the default decision threshold, while $\theta=0.80$ is a stricter test setting.}}
\label{tab:End-to-End-Fallback}
\vspace{-3mm}
    \begin{tabular}{lcccc}
    \toprule
    \textcolor{black}{Setting} & \textcolor{black}{threshold} & \textcolor{black}{Push Rate (\%)} & \textcolor{black}{Acc-S$_{\text{push}}$} & \textcolor{black}{SR$_{\text{push}}$} \\ 
    \midrule
    \textcolor{black}{w/o fallback} & \textcolor{black}{--} & \textcolor{black}{100.0} & \textcolor{black}{0.31} & \textcolor{black}{22.7} \\
    \multirow{2}{*}{\textcolor{black}{w/ confidence fallback}} & \textcolor{black}{0.80} & \textcolor{black}{14.0} & \textcolor{black}{0.44} & \textcolor{black}{34.6} \\
    & \textcolor{black}{0.60} & \textcolor{black}{40.0} & \textcolor{black}{0.39} & \textcolor{black}{29.4} \\
    \bottomrule
    \end{tabular}
\end{table}

\subsection{\textcolor{black}{Robustness Analysis}}

\subsubsection{Generalization Capability}

\textcolor{black}{To further examine the generalization of the Act2Intention Agent and Act2Intention Bench, we evaluated intent understanding and prediction on the OOD test set. The results in Table~\ref{tab:Generalization} show that, for both intent understanding and prediction tasks, the SFT-trained model outperforms the corresponding non-fine-tuned model on both ID and OOD test sets. However, a non-trivial OOD degradation remains. For example, in understanding, Llama-3.1-8B-SFT drops 12.73 Acc-G and 0.15 Acc-S; in prediction, Llama-3.1-8B-SFT declines by 10.65 Acc-C and 0.05 Acc-S.}

\textcolor{black}{A key observation is that while SFT improves ID performance for all models, it can also reduce generalization to OOD data. This leads to a more pronounced performance drop on OOD data compared to their non-fine-tuned counterparts and open-source models. For instance, in the intention understanding task, the OOD performance drop in Acc-G is larger for Deepseek-7B-SFT (-11.37) and smaller for Deepseek-7B (-0.25). Comparing these models, in intent understanding, Qwen-2.5-7B-SFT has the lightest degree of degradation (-7.50), while in intent prediction, Deepseek's decrease is relatively small (-7.67). Gemini-1.5-pro shows minimal performance fluctuation between the two sets. These findings suggest that future work can improve model generalization by selecting more suitable base models and incorporating online learning methods to continuously adapt user behavior patterns.}

\begin{table}[htbp]
\centering
\caption{Generalization evaluation. Red indicates the performance difference between the ID and OOD test sets.}
\label{tab:Generalization}

\begin{tabular}{l|ll|ll|ll|ll}
\toprule
\multirow{3}{*}{Method} &
\multicolumn{4}{c|}{Understanding} &
\multicolumn{4}{c}{Prediction} \\ \cmidrule{2-9}
& \multicolumn{2}{c|}{ID} & \multicolumn{2}{c|}{OOD} &
  \multicolumn{2}{c|}{ID} & \multicolumn{2}{c}{OOD} \\ \cmidrule{2-9}

& Acc-G & Acc-S & Acc-G & Acc-S & Acc-C & Acc-S & Acc-C & Acc-S \\ \midrule

Gemini-1.5-pro               &68.63 	&0.32 	&68.78 	\textcolor{red}{[+0.15]} 	&0.31 	\textcolor{red}{[-0.01]} 	&34.50  & 0.36  &	35.00 	\textcolor{red}{[+0.50]} 	    & 0.36 	\textcolor{red}{[-0.00]} \\ 
Llama-3.1-8B                 &6.03 	    &0.11 	&6.13 	\textcolor{red}{[+0.10]} 	&0.28 	\textcolor{red}{[+0.17]} 	&37.80  & 0.33  &	37.50 	\textcolor{red}{[-0.30]} 	    & 0.32 	\textcolor{red}{[-0.01]} \\ 
Llama-3.1-8B-SFT$\dagger$    &97.31 	&0.47 	&84.58 	\textcolor{red}{[-12.73]} 	&0.32 	\textcolor{red}{[-0.15]} 	&54.20  & 0.39  &	43.55 	\textcolor{red}{[-10.65]} 	    & 0.34 	\textcolor{red}{[-0.05]} \\  \midrule
Deepseek-7B                  &6.12 	    &0.19 	&5.87 	\textcolor{red}{[-0.25]} 	&0.19 	\textcolor{red}{[+0.00]} 	&46.89  & 0.34  &	40.93 	\textcolor{red}{[-5.96]} 	    & 0.28 	\textcolor{red}{[-0.06]} \\ 
Deepseek-7B-SFT$\dagger$     &100.00 	&0.50 	&88.63 	\textcolor{red}{[-11.37]} 	&0.37 	\textcolor{red}{[-0.13]} 	&53.07  & 0.42  &	45.40 	\textcolor{red}{[-7.67]} 	    & 0.35 	\textcolor{red}{[-0.07]} \\  \midrule
Qwen-2.5-7B                  &12.14 	&0.24 	&10.12 	\textcolor{red}{[-2.02]} 	&0.21 	\textcolor{red}{[-0.03]} 	&43.50  & 0.25  &	37.82 	\textcolor{red}{[-5.68]} 	    & 0.21 	\textcolor{red}{[-0.04]} \\ 
Qwen-2.5-7B-SFT$\dagger$     &100.00 	&0.49 	&92.50 	\textcolor{red}{[-7.50]} 	&0.45 	\textcolor{red}{[-0.04]} 	&50.00  & 0.40  &	39.40 	\textcolor{red}{[-10.60]} 	    & 0.37 	\textcolor{red}{[-0.03]} \\ 

\bottomrule

\end{tabular}
\end{table}

\subsubsection{Impact of Intention Length}
\begin{figure}[t]
\centering    
    \begin{minipage}[h]{0.48\linewidth}
        \centering
            \includegraphics[width=\linewidth]{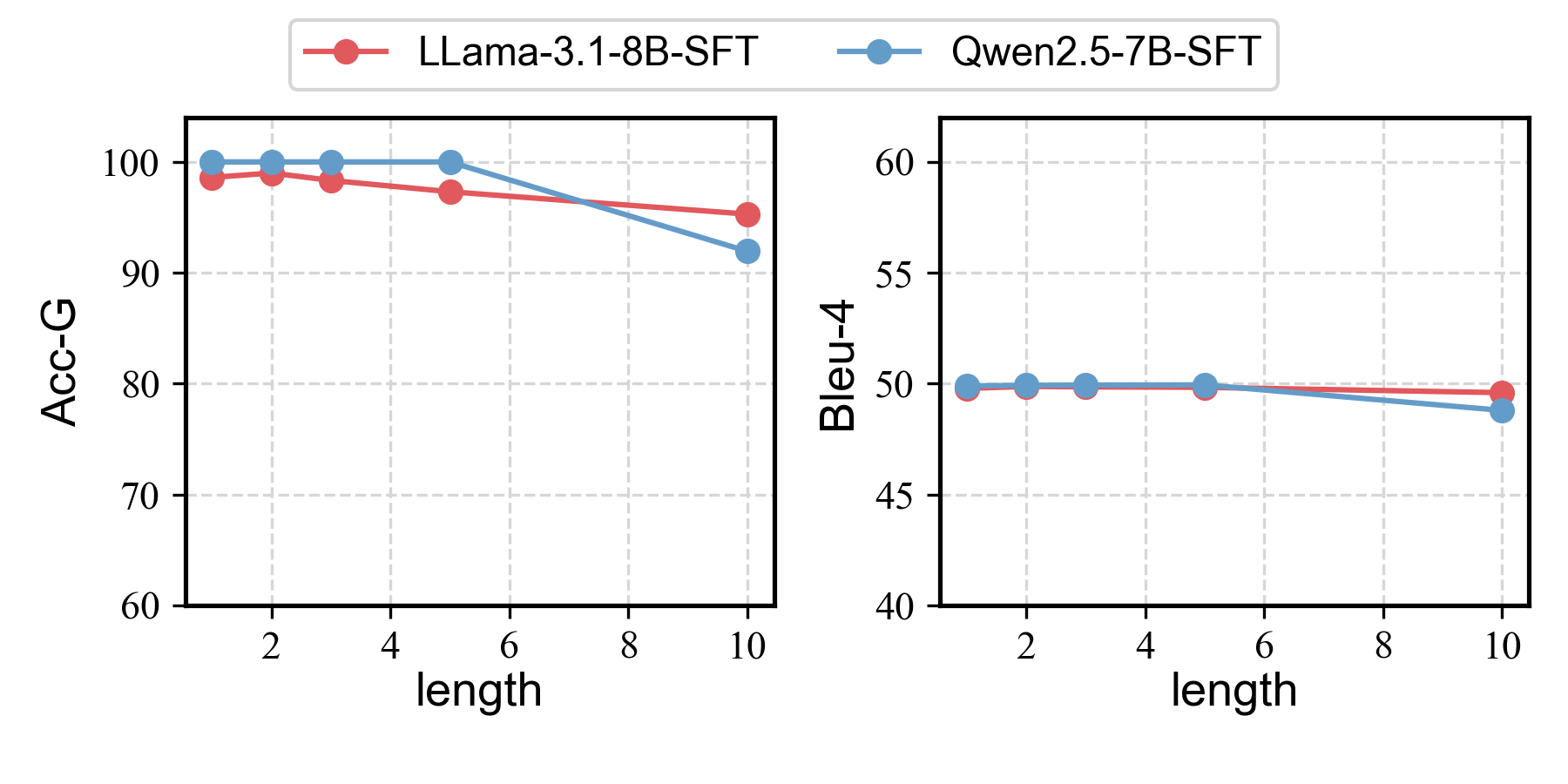}
            \subcaption{Understanding}\label{fig:length_understander}
    \end{minipage} 
    \begin{minipage}[h]{0.48\linewidth}
        \centering
            \includegraphics[width=\linewidth]{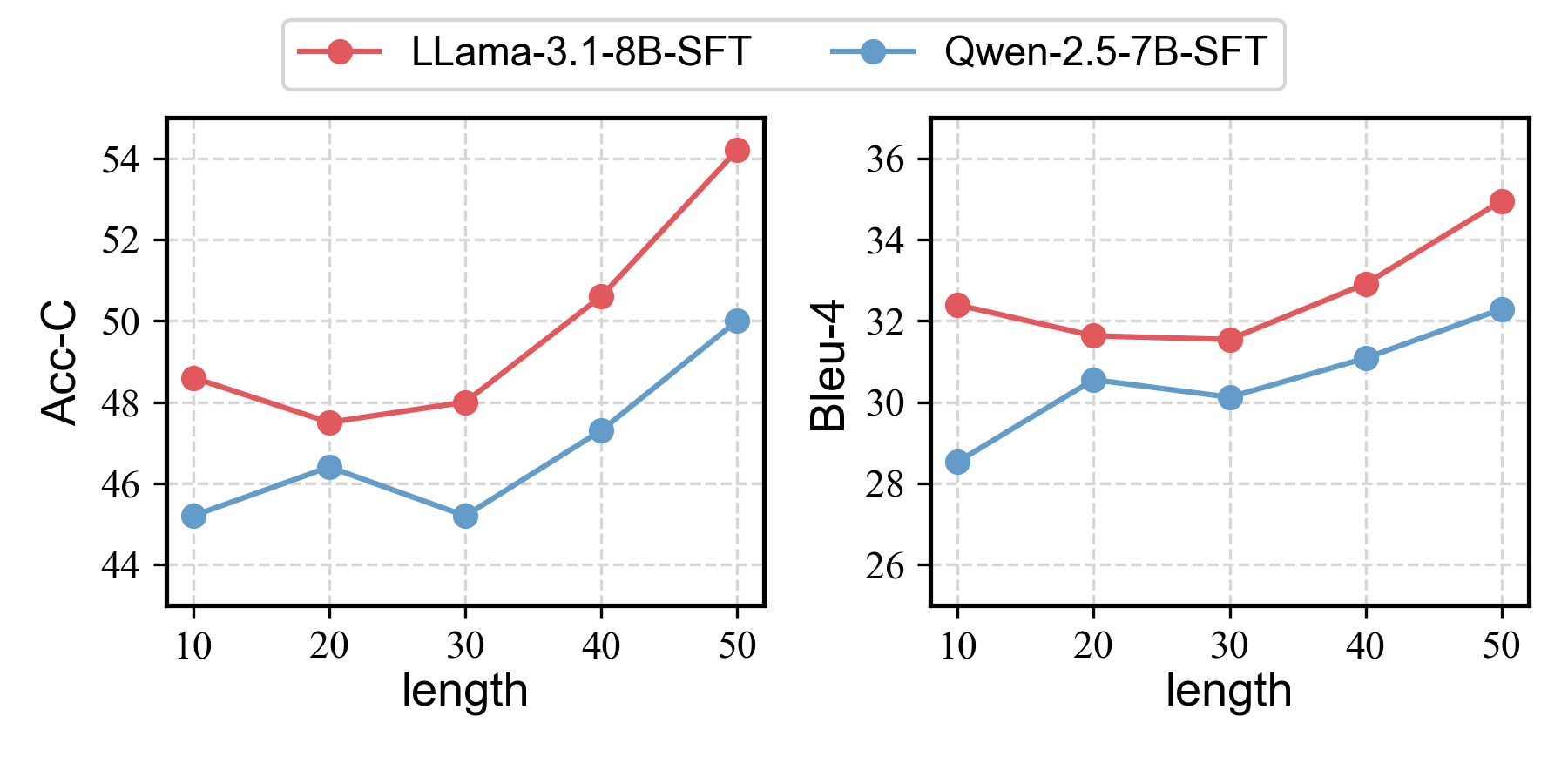}
        \subcaption{Prediction}\label{fig:length_Predictor}
    \end{minipage} 
\caption{Performance comparison of different trajectory sequence lengths. (a) represents the process of understanding segmentation accuracy and semantic accuracy; (b) represents the process of prediction category accuracy and semantic accuracy.}
  \label{fig:comparison_different_length}
\end{figure}

As shown in Figure~\ref{fig:comparison_different_length}, we studied how intention length impacts understanding or prediction performance. Specifically, for the intention understanding task, we fine-tuned the LLaMA-3.1-8B and Qwen-2.5-7B using sequences of length 5 and evaluated them on datasets with lengths of 1, 2, 3, 5, and 10. For the intention prediction task, we fine-tuned the same models using trajectories of length 50 and evaluated them on lengths of 10, 20, 30, 40, and 50.
\textcolor{black}{It is worth noting that the notion of continuity plays different roles in the two tasks. In intention understanding, longer action-description sequences increase the difficulty of segmentation and semantic grouping. In contrast, in intention prediction, longer historical intention trajectories provide richer behavioral context for anticipating the next intention. Therefore, the drop in long-sequence understanding does not contradict the value of continuous trajectories for proactive intention prediction.}

\textbf{Intention Understanding.}
As shown in Figure~\ref{fig:length_understander}, both LLaMA-3.1-8B-SFT and Qwen-2.5-7B-SFT maintain stable accuracy of at least 95\% for sequences of length 1–5. \textcolor{black}{This indicates that the fine-tuned models maintain high accuracy under moderate variations in action length in this experiment.} However, when the sequence length increases to 10, the group accuracy declines to 91.94\%. \textcolor{black}{This decline is likely due to the models being fine-tuned on sequences of length 5, so too-long inputs may exceed their optimal context window and reduce intention-understanding performance. Moreover, LLMs may exhibit Context Rot \cite{hong2025context} under long inputs, which can reduce their attention mechanism and information integration ability. This phenomenon may decrease attention to related actions in intent grouping tasks with a length of 10, resulting in a decline in performance.}

\textbf{Intention Prediction.}
As shown in Figure~\ref{fig:length_Predictor}, both models exhibit a consistent upward trend as the input length increases from 10 to 50. This is expected because the models were trained on sequences of length 50. From length 10 to 40, LLaMA-3.1-8B-SFT improves from 48.6 to 50.6 Acc-C and from 32.4 to 32.9 BLEU-4, while Qwen-2.5-7B-SFT increases from 45.2 to 47.3 Acc-C and from 28.5 to 31.8 BLEU-4. This indicates that longer intention histories enhance temporal reasoning and sequence coherence, allowing the models to capture user intentions more accurately.

\subsubsection{Impact of Training Data} 

In order to dissect the contribution of real versus generated data in our benchmark, we fine-tuned the Qwen-2.5-7B for intention understanding and prediction across four distinct training sets: RR (Real-persona-to-Real-trajectory), RG (Real-persona-to-Generated-trajectory), GG (Generated-persona-to-Generated-trajectory), and RR+RG+GG. 

\begin{figure}[htbp]
	\centering
	\begin{minipage}[c]{0.48\textwidth}
		\centering
		\includegraphics[width=\textwidth]{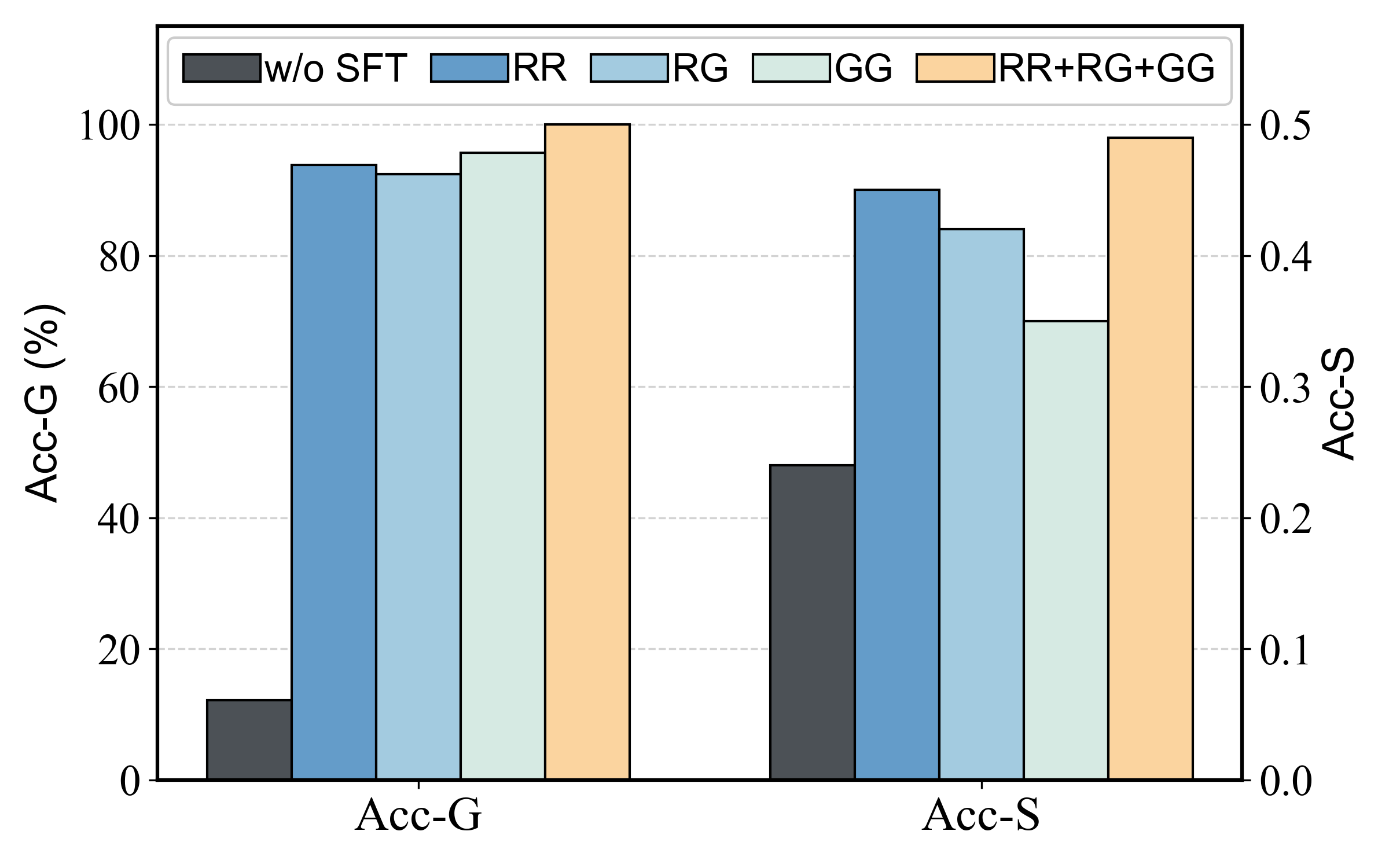}
		\subcaption{Understanding}\label{fig:Robust_Generated_data_Understanding}
	\end{minipage} 
	\begin{minipage}[c]{0.48\textwidth}
		\centering
		\includegraphics[width=\textwidth]{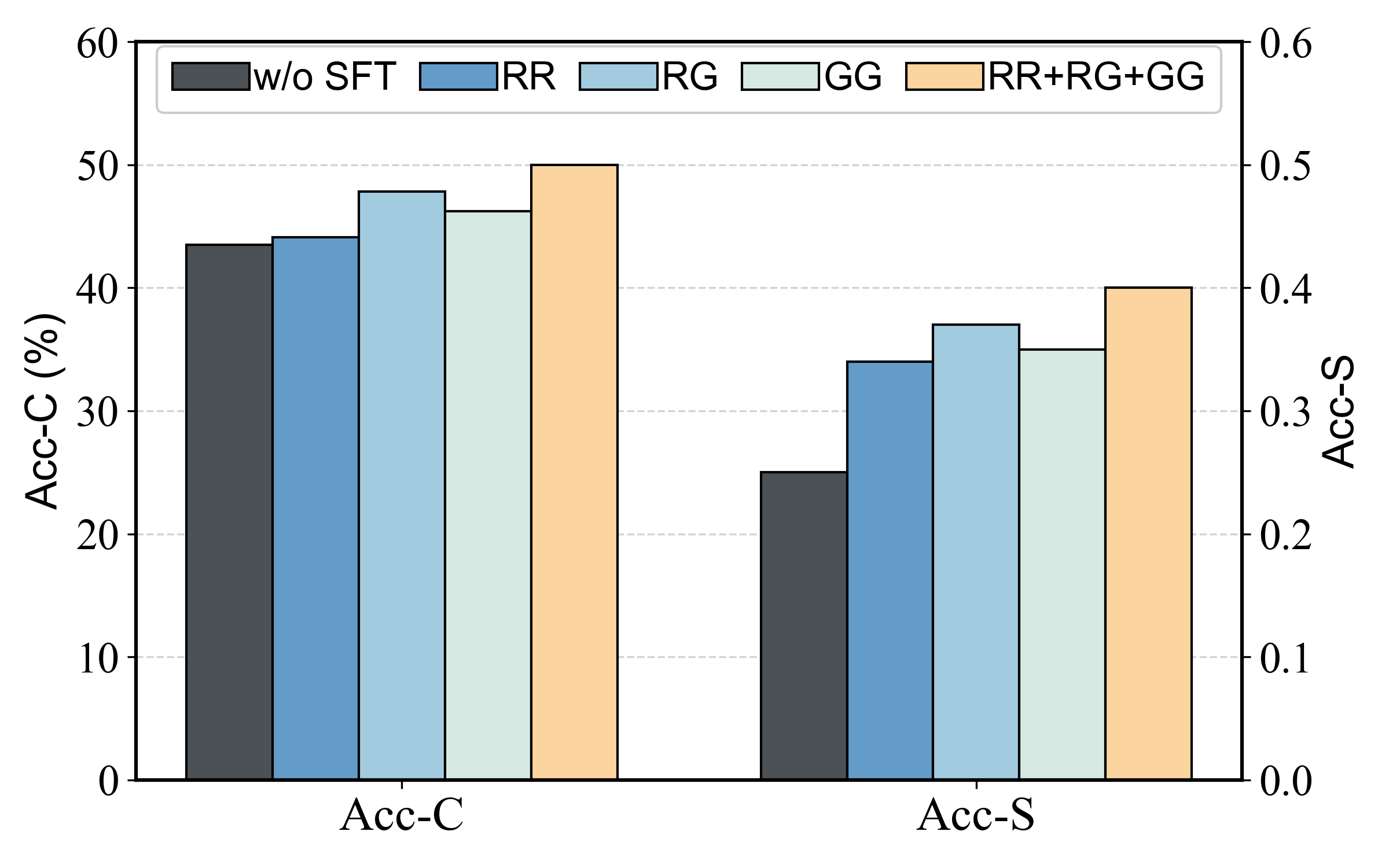}
		\subcaption{Prediction}\label{fig:Robust_Generated_data_Prediction}
	\end{minipage}
    \caption{\textcolor{black}{Performance of different training sets. We fine-tuned Qwen-2.5-7B; (a) shows the results on intention understanding, and (b) shows the results on intention prediction. The combination ``RR+RG+GG'' achieves the highest overall performance in this experiment, suggesting that real and generated data are complementary for improving model generalization.}}
    \label{fig:Robust_Generated_data}
\end{figure}

As shown in Figure~\ref{fig:Robust_Generated_data}, all training sets contribute positively to model performance, while their combination yields the best results. For the intention understanding task, compared with RR and RG, GG achieves the highest Acc-G (95.7) but the lowest Acc-S (0.35). This is partly because GG has a larger amount of data, which supports better grouping of intentions. On the other hand, purely synthetic data cannot provide complete semantic information for the model to learn, leading to a gradual decrease in Acc-S from RR to RG to GG. When all subsets are combined (RR+RG+GG), the model achieves perfect classification accuracy (100.0) and the highest semantic consistency (Acc-S = 0.49), showing strong complementarity among data.

\subsubsection{Impact of Persona}

\begin{figure}[htbp]
	\centering
	\begin{minipage}[c]{0.48\textwidth}
		\centering
		\includegraphics[width=\textwidth]{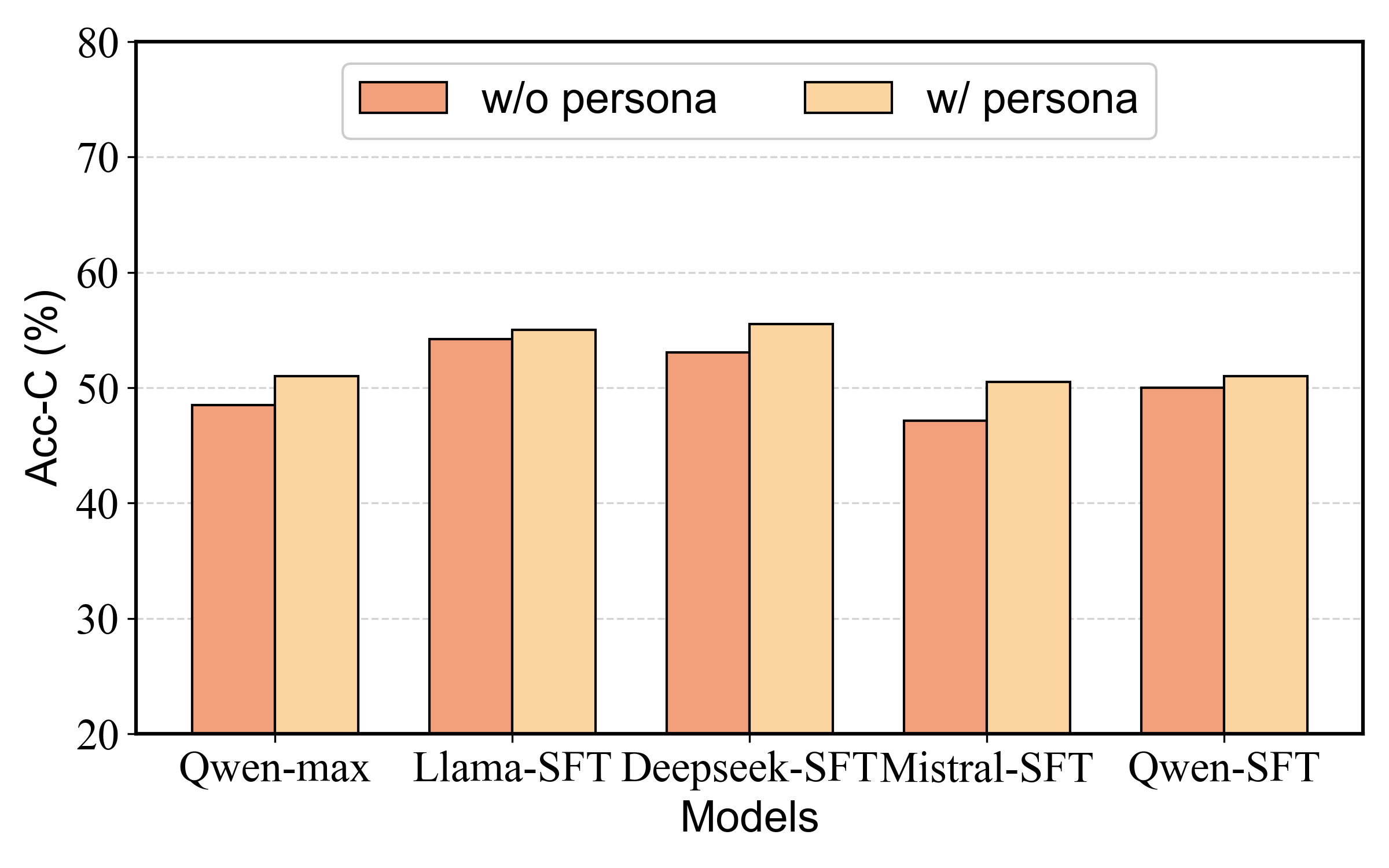}
		\subcaption{Acc-C}\label{fig:robust_persona_Acc_C}
	\end{minipage} 
	\begin{minipage}[c]{0.48\textwidth}
		\centering
		\includegraphics[width=\textwidth]{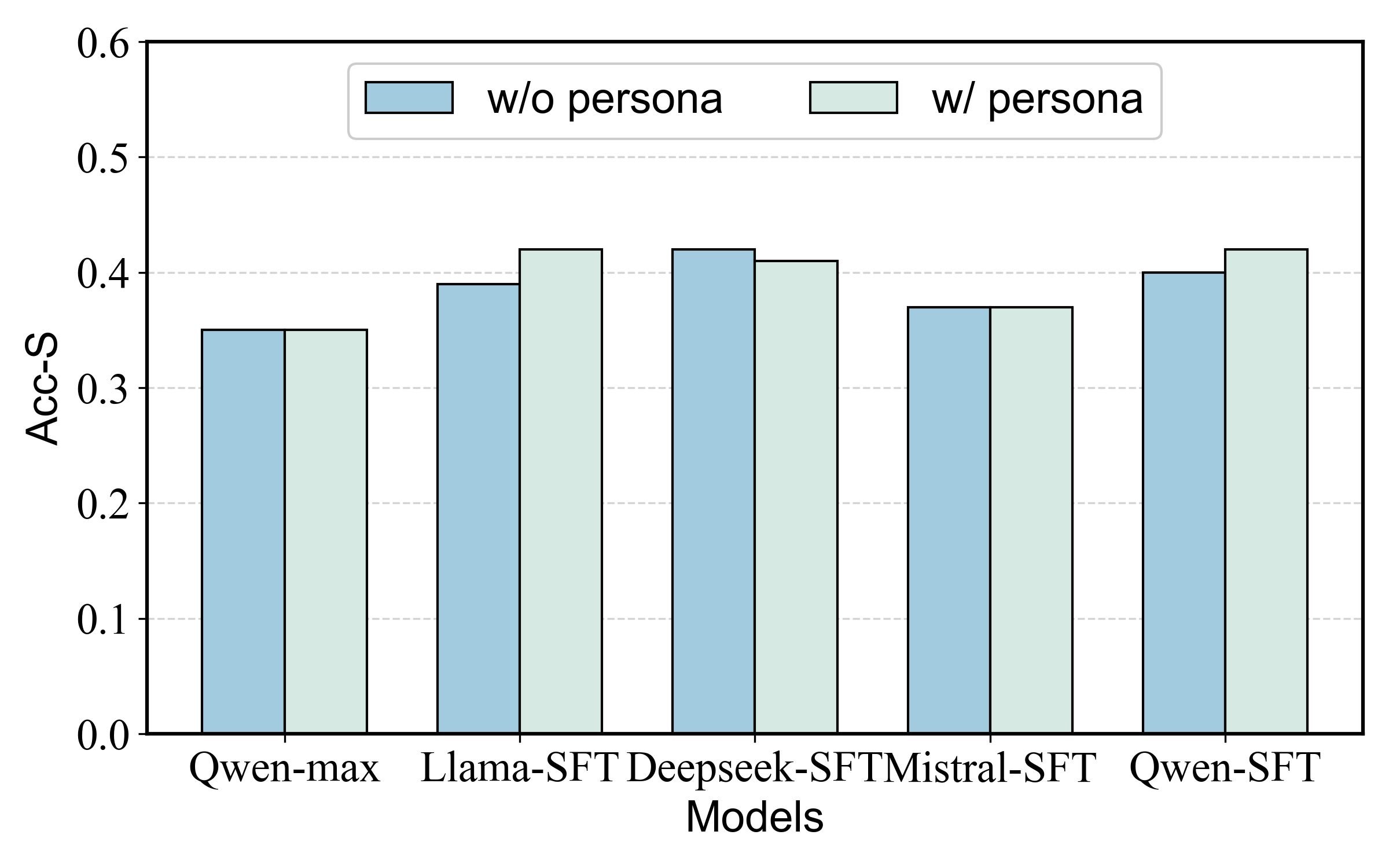}
		\subcaption{Acc-S}\label{fig:robust_persona_Acc_S}
	\end{minipage}
    \caption{Effect of Persona on Intention Prediction. }
    \label{fig:Robust_Persona}
\end{figure}

To verify the impact of user persona on intention prediction performance, we conduct an ablation study across several LLMs. Specifically, we compare the performance of Qwen-max, Llama-3.1-8B-SFT, Deepseek-7B-SFT, Mistral-7B-SFT, and Qwen-2.5-7B-SFT when inferring intention with and without persona in the input context, respectively. Here, all SFT models were trained without persona information. As shown in Figure~\ref{fig:robust_persona_Acc_C}, for the category accuracy (Acc-C), the metrics of all models improved when the persona is used. For instance, Deepseek-7B-SFT shows an increase from 53.07\% to 55.5\%. This indicates that information such as user preferences in the persona helps the model narrow down the category of the user's next intention. However, as shown in Figure~\ref{fig:robust_persona_Acc_S}, the impact on semantic accuracy (Acc-S)  is less consistent. LLaMA-3.1-8B-SFT and Qwen-2.5-7B-SFT show slight improvements (+0.02 and +0.01), while Mistral-7B-SFT, Qwen-max, and Deepseek-7B-SFT remain nearly unchanged or degraded. This indicates that while persona information enhances the model’s ability to identify the correct intention type, it does not substantially improve the semantic details between predicted and reference intentions. For example, personas mainly influence high-level preference reasoning (e.g., ``user prefers to open shopping apps'') rather than detailed information (e.g., ``user prefers to buy women's jeans''). \textcolor{black}{Overall, these findings suggest that persona knowledge mainly supports categorical user-goal understanding; however, noisy, incomplete, outdated, or inaccurate personas may still bias downstream intention prediction, so future work should study robust and privacy-preserving persona updating.}

\vspace{-5mm}

\section{CASE STUDY}
\begin{figure*}[htb]
  \centering
  \includegraphics[width=0.55\linewidth]{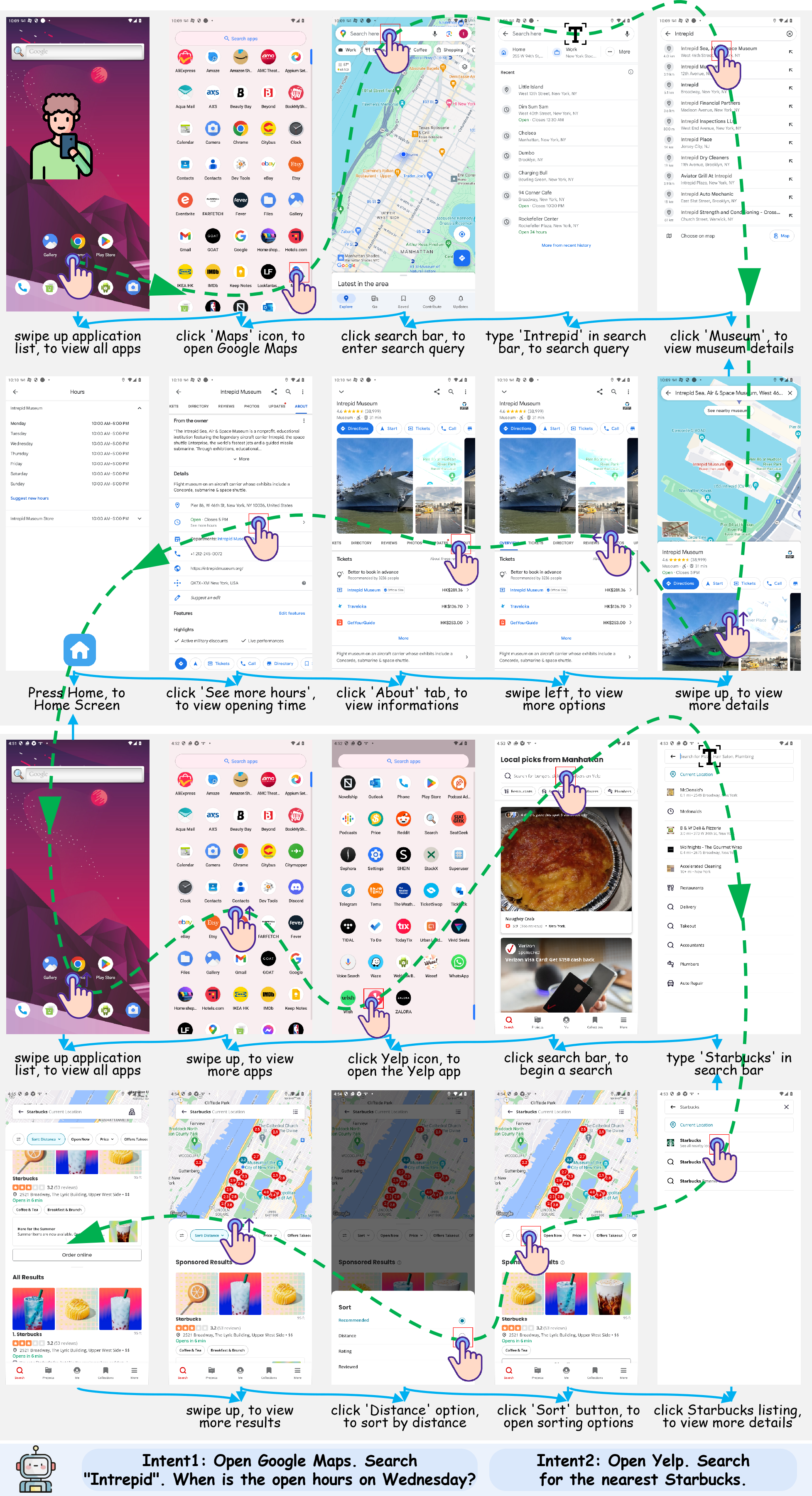}
  \caption{A case of Intention Understanding and Prediction.}
  \label{fig:case_1}
\end{figure*}

Figure~\ref{fig:case_1} illustrates the operational process of Intention Understanding. For each action, the agent generates a natural language description of both the action and its purpose by analyzing the split-screenshots before and after its execution. Building upon these action descriptions, the agent can then effectively infer the underlying intentions corresponding to the sequence of actions.

\end{document}